\documentclass[sigconf]{acmart}
\AtBeginDocument{%
  \providecommand\BibTeX{{%
    \normalfont B\kern-0.5em{\scshape i\kern-0.25em b}\kern-0.8em\TeX}}}

\copyrightyear{2021} 
\acmYear{2021} 
\setcopyright{acmcopyright}\acmConference[SIGIR '21]{Proceedings of the 44th International ACM SIGIR Conference on Research and Development in Information Retrieval}{July 11--15, 2021}{Virtual Event, Canada}
\acmBooktitle{Proceedings of the 44th International ACM SIGIR Conference on Research and Development in Information Retrieval (SIGIR '21), July 11--15, 2021, Virtual Event, Canada}
\acmPrice{15.00}
\acmDOI{10.1145/3404835.3462833}
\acmISBN{978-1-4503-8037-9/21/07}

\usepackage{amsmath,amsfonts,amsthm,bm} 
\usepackage{natbib}
\usepackage{tikz}
\usepackage{pgfplots}
\pgfplotsset{compat=1.16}
\usetikzlibrary{patterns}
\usepackage{subcaption}
\usepackage[font=small]{caption}
\usepackage{multirow}
\usepackage{array}
\newcolumntype{P}[1]{>{\centering\arraybackslash}p{#1}}
\usepackage{balance}
\usepackage{hyperref}

\usepackage{enumitem}

\usepackage{amsfonts}

\def\-{\mbox{-}}

\def\la{\langle}

\def\ra{\rangle}

\def\Ju{\mathcal{J}^U}
\def\Jz{\mathcal{J}^I}

\def\*{\star}

\newcommand{\argmin}{arg\,min}

\usepackage{xcolor}

\definecolor{Mulberry}{rgb}{0.77,0.29,0.55}
\definecolor{CadmiumOrange}{rgb}{0.93,0.53, 0.18}
\definecolor{ForestGreen}{rgb}{0.13, 0.55, 0.13}
\definecolor{WildStrawberry}{rgb}{0.5, 0.7, 0.2}

\newcommand{\norm}[1]{\left\lVert#1\right\rVert}

\pgfplotsset{mystyle2/.style={%
        font=\rmfamily\Labelsize,
        width=0.225\textwidth,
        mark size=1.1pt,
        ylabel near ticks,
        xlabel near ticks,
        label style = {font=\small},
        tick label style = {font=\tiny, yshift=0.5ex},
        ylabel shift = -5 pt, 
        xlabel shift = -5 pt,
        title style={yshift=-1.2ex,font=\small},
        y tick label style={/pgf/number format/.cd,fixed,fixed zerofill,precision=2,/tikz/.cd},
        legend image post style={scale=0.5},
        every axis plot/.append style={semithick},
        legend style={font=\scriptsize, mark size=4pt},
        legend columns=1, legend style={/tikz/column 2/.style={column sep=0.5pt}},
        mark options={scale=1.7}}}

\pgfplotsset{xbarstyle/.style={%
    xbar,
    width=0.225\textwidth,
    label style={font=\small},
    tick label style={font=\tiny},
    axis x line*=bottom,
    axis y line*=none,
    height=3.7cm,
    title style={yshift=-0.5ex,font=\small},
    ylabel shift = -5 pt, 
    xlabel shift = -5 pt,
    ylabel near ticks,
    xlabel near ticks,
    tickwidth         = 3pt,
    ytick=data,
    symbolic y coords = {<\textit{MLP, None}>, \textbf{<\textit{MLP, Bi}>}, <\textit{MLP1, MLP2}>, <\textit{Bi, MLP}>, <\textit{Bi, Bi}>, <\textit{MLP, MLP}>},
    bar width=1mm,
    xlabel style={font=\footnotesize}
}}

\pgfplotsset{barstyle/.style={%
    ybar,
    width=0.225\textwidth,
    label style={font=\small},
    tick label style={font=\tiny,yshift=0.5ex},
    axis x line*=bottom,
    axis y line*=none,
    title style={yshift=-0.5ex,font=\small},
    ylabel shift = -5 pt, 
    xlabel shift = -5 pt,
    ylabel near ticks,
    xlabel near ticks,
    tickwidth         = 3pt,
    ytick=data,
    bar width=1mm,
    symbolic x coords = {None, User, Item, Both},
    legend style={font=\scriptsize, mark size=4pt},
    y tick label style={/pgf/number format/.cd,fixed,fixed zerofill,precision=2,/tikz/.cd},
    legend columns=1, legend style={/tikz/column 2/.style={column sep=0.5pt}},
    label style = {font=\small}
}}

\definecolor{mycolor1}{RGB}{31, 95, 139}
\definecolor{mycolor2}{RGB}{122, 81, 149}
\definecolor{mycolor3}{RGB}{239, 86, 117}
\definecolor{mycolor4}{RGB}{255, 166, 0}

\begin{document}
\fancyhead{}

\title{Neural Graph Matching based Collaborative Filtering}

\author{Yixin Su}
\affiliation{%
\institution{University of Melbourne\country{Australia}}
}
\email{yixins1@student.unimelb.edu.au}
\author{Rui Zhang}
\authornote{Corresponding authors.}
\authornote{Rui Zhang did this work while working at the University of Melbourne.}
\affiliation{%
\institution{\href{https://www.ruizhang.info/}{www.ruizhang.info}\country{China}}
}
\email{rayteam@yeah.net}

\author{Sarah M. Erfani}
\authornotemark[1]
\affiliation{%
\institution{University of Melbourne\country{Australia}}
}
\email{sarah.erfani@unimelb.edu.au}
\author{Junhao Gan}
\affiliation{%
\institution{University of Melbourne\country{Australia}}
}
\email{junhao.gan@unimelb.edu.au}

\begin{abstract}
User and item attributes are essential side-information; their interactions (i.e., their co-occurrence in the sample data) can significantly enhance prediction accuracy in various recommender systems. We identify two different types of attribute interactions, \textit{inner interactions} and \textit{cross interactions}: inner interactions are those between \textit{only} user attributes or those between \textit{only} item attributes; cross interactions are those between user attributes and item attributes. Existing models do not distinguish these two types of attribute interactions, which may not be the most effective way to exploit the information carried by the interactions. To address this drawback, we propose a neural \underline{G}raph \underline{M}atching based \underline{C}ollaborative \underline{F}iltering model (GMCF), which effectively captures the two types of attribute interactions through modeling and aggregating attribute interactions in a graph matching structure for recommendation. In our model, the two essential recommendation procedures, characteristic learning and preference matching, are explicitly conducted through graph learning (based on inner interactions) and node matching (based on cross interactions), respectively. Experimental results show that our model outperforms state-of-the-art models. 
Further studies verify the effectiveness of GMCF in improving the accuracy of recommendation.
\end{abstract}


\maketitle

\section{Introduction}
Collaborative Filtering (CF) is one of the most frequently used algorithms in recommender systems. It performs the predictions based on an assumption that similar users will have common preferences on similar items. For example, matrix factorization based recommender systems \cite{koren2009matrix,he2017neuralCF, wang2021combating} learn user and item embeddings from user-item interactions (e.g., click, purchase) to exhibit similarities between similar users and between similar items. 
One of the most important challenges faced by CF is how to consider user and item attributes (e.g., user genders, item colors) to enhance the prediction performance. 
To this end, recent CF models conduct attribute embedding  to capture fine-grained collaborative information and reveal the similarities between attributes \cite{adams2010incorporating,su2019mmf}.
While learning attribute embeddings, considering the co-occurrence of attributes inside each data sample, i.e., \textit{attribute interactions}, have been proven essential in providing useful information for more accurate predictions \cite{rendle2010factorization,song2019autoint}. 
For example, in a movie recommendation system, the director \textit{Nolan} is a master of producing \textit{sci-fi} movies. In this scenario, considering the attribute interaction $<$\textit{Nolan}, \textit{sci-fi}$>$ is more effective than considering the two attributes separately. 
Generally, an attribute-interaction-aware CF model takes the attribute interactions into account (by modeling and aggregating them) to jointly decide the final predictions.
Factorization Machine (FM) \cite{rendle2010factorization} models each attribute interaction as a dot product of two embedded vectors and aggregates all the modeling results linearly. 
With the development of Graph Neural Networks (GNNs), \citet{li2019fi} and \citet{su2020detecting} leverage the relation modeling ability of GNNs to capture more sophisticated attribute interaction information and aggregate the information through graph learning.

While these models capture the co-occurrence between attributes by modeling attribute interactions, we argue that they may not be sufficient to yield a satisfactory joint decision.
The key reason is that these models simply treat all the attribute interactions equally, and hence, model and aggregate them in the same way.
However, different attribute interactions should have different impacts on the final prediction.
Specifically, when the attribute interactions are exploited to perform joint decisions, 
they could play different roles for recommendation. 
For example, without loss of generality, the goal of movie recommendation is to predict the preference of a user (represented by the user's attributes) on a movie (represented by the item's attributes). 
Therefore, the attribute interactions are naturally divided into two categories based on their roles for recommendation. 
First, the interactions between user attributes {\em only} (or between item attributes {\em only}), 
e.g., $<$\textit{male}, \textit{18-24}$>$, $<$\textit{Nolan}, \textit{sci-fi}$>$, serve as user (item) characteristic learning. 
These interactions are called \textit{inner interactions}.
Second, the interactions between user attributes and item attributes, e.g., $<$\textit{18-24}, \textit{sci-fi}$>$, serve as preference matching, such as ``whether users in 18-24 years old like sci-fi movies''. 
These interactions are called \textit{cross interactions}.
Existing work do not distinguish these two types of interactions. 
Such unawareness of attribute interaction types inevitably limits the capability of the existing work of leveraging attribute interactions for producing satisfactory joint decisions. 

\begin{figure}[t]
\centering
\centerline{\includegraphics[width=0.88\columnwidth]{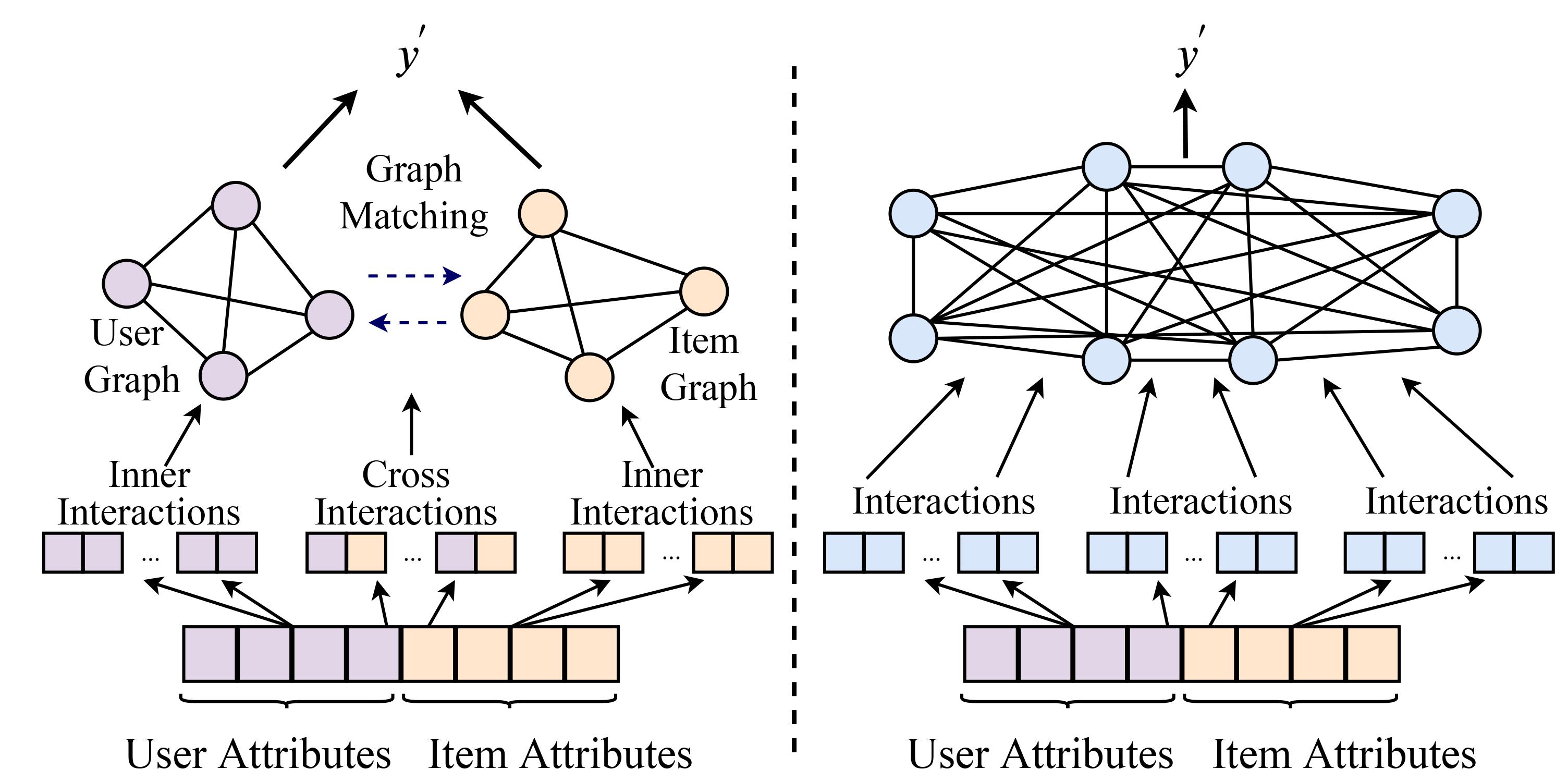}}
\vskip -0.08in
\caption{Illustration of the differences between GMCF (left) and existing work (right). GMCF treats attribute interactions differently in a structure of graph matching, while existing work treats all attribute interactions equally.}
\label{fig:running_example}
\vskip -0.15in
\end{figure}

In this work, we explicitly model and aggregate inner interactions and cross interactions in different ways in a graph matching structure. 
Specifically, we use a graph composed of a user's (an item's) attributes to represent the user (the item). 
Each attribute is a node, and each pairwise attribute interaction is an edge. 
Then, we propose a neural Graph Matching based Collaborative Filtering model, called GMCF.
Our GMCF uses a GNN to model a user attribute graph and an item attribute graph based on their inner interactions, respectively.
Meanwhile, it matches the two attribute graphs at node level based on cross interactions, and makes the final prediction result.
Figure \ref{fig:running_example} illustrates the differences between GMCF and existing work (GNN-based work as an example) in modeling and aggregating attribute interactions. 
As a result, GMCF consider attribute interactions in a way that fits what a recommender system aims to do: learn user and item characteristics (through graph learning), and match users' preference to items based on their characteristics (through graph matching).

We summarize the main contributions of our work as follows:
\begin{itemize}[leftmargin = *]
    \item We highlight the importance of considering the different impacts of attribute interactions on recommendation predictions, and categorize them into inner interactions and cross interactions based on their roles for recommendation, which deserve to be treated differently when they jointly decide the predictions.
    \item We propose a novel neural graph matching-based model, GMCF, that effectively leverages inner interactions for user and item characteristic learning (through graph learning), and exploits cross interactions for preference matching (through graph matching). 
    \footnote{Our implementation of the GMCF model is available at: \href{https://github.com/ruizhang-ai/GMCF\_Neural\_Graph\_Matching\_based\_Collaborative\_Filtering}{https://github.com/ruizhang-ai/GMCF\_Neural\_Graph\_Matching\_based\_Collaborative\_Filtering}.} 
    To our best knowledge, we are the first to represent each user and each item as an attribute graph, and model and aggregate attribute interactions in a graph matching structure.
    \item We conduct extensive experiments. Experimental results show that (i) our model outperforms state-of-the-art baselines in terms of accuracy; (ii) GMCF is able to effectively model the two types of attribute interactions to produce accurate predictions.
\end{itemize}

\section{Related Work}
\label{sec:related_work}
In this section, we discuss more related work on attribute-aware CF models, graph neural networks, and graph matching methods.

\subsection{Attribute-aware CF models}
Factorization machine (FM) \cite{rendle2010factorization} is one of the most frequently used collaborative filtering algorithms that takes attribute interactions into account. FM performs interaction modeling between each pair of attributes and sums up all the modeling results to make the final prediction.
Some extensions of FM further calculates an attention score \cite{xiao2017attentional} or a gate \cite{liu2020autofis} for each interaction modeling result to differentiate each interaction's importance using the attention mechanism. However, these models consider the structural information linearly, which is ineffective in leveraging attribute interaction to make a joint decision.
NFM \cite{he2017neural} and DeepFM \cite{guo2017deepfm} add a multilayer perceptron (MLP) on top of attributes or attribute interactions, aiming to implicitly capture the structural information in a non-linear way.
However, the use of MLP to model the interactions between input variables has been proven less effective \cite{beutel2018latent,song2019autoint}.
GMCF models and aggregates attribute interactions explicitly in a structure of graph matching, which is more effective in attribute interaction modeling and structural information capturing.  

\subsection{Graph Neural Networks}
Graph neural networks (GNNs) facilitate learning about entities and their relations \cite{kipf2016semi,xu2018powerful,zhao2021wgcn,pang2021graph}. Existing work leverages GNNs in various domains, such as molecular property prediction \cite{duvenaud2015convolutional,gilmer2017neural} and object relation learning physical systems \cite{chang2016compositional,battaglia2016interaction}.
Recently, GNNs attract attention in the domain of recommender systems. Some work considers the user-item interactions as a bipartite graph, where an edge between a user and an item indicates an interaction (e.g., click or rate) \cite{berg2017graph,huang2019carl,wang2019neural}. These models only consider user-item interactions in GNNs.
Other work leverages GNNs to model knowledge graphs for recommendation \cite{zhang2016collaborative,xian2019reinforcement,wang2019knowledge,wang2019kgat}. These models regard edges as predefined relations between attributes and items (users) instead of between attributes. Therefore, they do not consider the attribute interactions.
\citet{li2019fi} and \citet{su2020detecting} leverage GNNs to perform attribute interaction modeling and aggregation as a graph learning procedure.
However, these models analyze all attribute interactions equally, which are ineffective in capturing useful structural information of attribute interactions to make a joint decision.
Our model differentiates inner interactions and cross interactions, and models and aggregates these interactions in a graph matching structure that is considered as more suitable for recommendation.

\subsection{Graph Matching}
Graph matching is a long-standing research topic in computer science such as database and data mining domains \cite{yan2005substructure, dijkman2009graph}. The goal of graph matching is to discover the similarity between two graph form representations.  
Traditional graph matching algorithms are based on heuristic rules, such as minimal graph edit distance \cite{willett1998chemical, raymond2002rascal}, or based on graph kernel methods, such as random walks inside graphs \cite{kashima2003marginalized, vishwanathan2010graph} and graph sub-structures \cite{shervashidze2009efficient, shervashidze2009fast}. 
With the development of GNNs, recent work explores neural-based graph matching. \citet{bai2019simgnn} fuses the graph-level embeddings learned by GNNs and a node-level matching embedding learned by a pairwise node comparison method. \citet{li2019graph} uses GNNs to learn two graphs' embeddings in a Siamese network framework \cite{bromley1994signature}, with a node-level matching performed in the fusing procedure.   
However, the neural graph matching methods have not been fully explored in recommender systems yet. 
We are the first to represent each user and each item in a graph form and leverage the framework of neural graph matching for preference matching.

\section{Problem Statement \& Background}

In this section, we introduce the problem definition, the representations of attribute interactions, and the basic idea of GNNs.

\subsection{Problem Statement}
Denote by $\Ju$ the {\em universe} of the user attributes and by $\Jz$ the {\em universe} of the item attributes.
An {\em attribute-value pair} is a name-value pair, denoted by $(att, val)$, where $att$ is the name of the attribute and $val$ is the value on this attribute.
For example, $(Male, 1)$ and $(Female, 1)$ mean the user's gender is Male and Female, respectively, where $Male \in \Ju$ and $Female \in \Ju$ are considered as two user attributes.
Let $D$ be a set of $N$ training data pairs, denoted by $D=\{(\bm{x}_{n}, y_{n})\}_{1\leq n \leq N}$.
In each training data $(\bm{x}_n, y_n) \in D$,
\begin{itemize}[leftmargin = *]
	\item $\bm{x}_n$ is called a {\em data sample}, 
		which consists of:
		\begin{itemize}[leftmargin = *]
			\item a set $C_n^U = \{c^U = (att, val)\}_{att\in J^U_n}$ of attribute-value pairs with respect to a set of user attributes $J^U_n \subseteq \Ju$, and 
			\item a set $C_n^I = \{c^I = (att, val)\}_{att\in J^I_n}$ of attribute-value pairs with respect to a set of item attributes $J^I_n \subseteq \Jz$.
		\end{itemize}
		Moreover, $C_n^U$ and $C_n^I$ are respectively called the {\em characteristic} of the user and the item specified in this data sample.
	\item $y_n \in \mathbb{R}$ is the {\em implicit feedback} (e.g., watched, liked) of the user on the item. 
\end{itemize}
It should be noted that the number of attribute-value pairs in each data sample may be different, as the information of some attributes may be missing, or may contain {\em multiple} attributes of the same type, e.g., a movie may belong to multiple genres.

\vspace{1mm}
\noindent
{\bf The Recommendation Problem.}
The goal of the {\em Recommendation Problem} studied in this paper is to design a {\em predictive model} $F(\bm{x}_n)$ such that for an input data sample  $\bm{x}_n$ (specifying the characteristics of a user and an item), the output of the model, $F(\bm{x}_n)$, is a prediction on the {\em true} feedback, $y_n$,  of the user on the item.   

\vspace{1mm}
\noindent
{\bf Our Solution.}
In this paper, we propose a {\em Neural \underline{G}raph \underline{M}atching based \underline{C}ollaborative \underline{F}iltering} (GMCF) model $F_{GMCF}(\bm{x}_n)$. Our GMCF models the user and the item characteristics as two graphs, respectively, and leverages these user and item graph representations to predict $y_n$ by graph matching techniques.

\subsection{Representing Attributes and Interactions}
	Each attribute $att \in \Ju \cup \Jz$ is embedded as a vector $\bm{v}$ in $d$-dimensional space $\mathbb{R}^d$.
	This process can be seen as building a parameter matrix as an embedding lookup table.
	All the data samples share the same vector embedding $\bm{v}_{att}$ of the same attribute $att$, but they may have different {\em scalar} on the vector due to
	the potentially different values $val$'s.
	More specifically, for an attribute-value pair $(att, val)$, the corresponding vector, $\bm{u}_{att}$, is computed as $\bm{u}_{att} = val \cdot \bm{v}_{att}$;
	and such $\bm{u}_{att}$ is called the {\em representation} of the attribute-value pair.
	Initially, the $\bm{v}_{att}$ of each attribute is set as a random vector.	
	In the following, when the context of the attribute $att$ is clear, we omit the subscript from $\bm{v}$ and $\bm{u}$ for simplicity.

	The co-occurence of two attributes $att_1$ and $att_2$ in a data sample is defined as an {\em interaction} between $att_1$ and $att_2$.
	Such an interaction is modeled by a function $f(\bm{u}_1, \bm{u_2}): \mathbb{R}^{2\times d} \rightarrow \mathbb{R}^\ell$, where
	$\bm{u}_1$ and $\bm{u}_2$ are the representations of the attribute-value pairs of $att_1$ and $att_2$ (in the same data sample), and $\ell$ is the dimensionality of the output.
	Since the set of attributes, $J^U_n$ and $J^I_n$, in each data sample may be different, the interactions specified in different data samples could be different.
	And because each attribute may appear (with different values) in multiple data samples, 
	the collaborative information of the interactions in different data samples 
	would further help discover the interactions between attributes that have never co-occurred. 
Therefore, $f(\cdot, \cdot)$ actually learns attribute embeddings that capture the collaborative information between the attributes 
(i.e., similar attributes would have similar embeddings) \cite{rendle2010factorization}.

Our proposed model GMCF categorizes attribute interactions in a data sample into two types, {\em inner interactions} and {\em cross interactions}. 
More specifically, the interactions between user attributes only and between item attributes only are defined as inner interactions.
However, on the other hand, those between one user attribute and one item attribute are cross interactions.
As we will see shortly in Section \ref{sec:our_approach}, our GMCF model deploys different functions, $f(\cdot, \cdot)$, to model these two kinds of attribute interactions, and respectively uses them for different purposes: (i) user and item characteristic learning, and (ii) recommendation.

\subsection{Graph Neural Networks}

Consider a graph $G= \la V, E \ra$, where 
$V = \{\bm{v}_i\}_{1 \leq i \leq k}$ is the set of $k$ nodes, each of which is represented by a vector representation $\bm{v}_i$, and 
$E$ is the set of edges which indicate the neighborhood information between nodes: two nodes are neighbors if they are linked by an edge.
A Graph Neural Network (GNN) learns the vector representation of each node by message passing, a procedure of aggregating neighborhood information.
Specifically, the message passing procedure for node $i$ first aggregates the vector representations of all its neighbors. 
Then, it gets the fused representation of node $i$ by fusing $\bm{v}_i$ and the aggregated vector representation. 
Formally, the fused vector representation of node $i$, $\bm{v}^{'}_{i}$, through the graph modeling is:
\begin{equation}
\small
\label{fun:node_update}
    \bm{v}^{'}_{i} = f_{fuse}(\bm{v}_i, Aggregate_{v}(\{\bm{v}_j\}_{j\in N(i)})),
\end{equation}
where $f_{fuse}$ is the fusing function, $Aggregate_{v}(\cdot)$ is an aggregation function that aggregates the neighborhood embeddings into a fixed dimension representation (e.g., element-wise sum) and $N(i)$ is the set of all the neighbors of node $i$.

If it is necessary, 
the graph representation can be computed as the aggregation of the vector embeddings of all the nodes: $\bm{v}_G = Aggregate_{G}(\{\bm{v}_i^{'}\}_{i\in V})$, 
where $\bm{v}_G$ is the graph representation and $Aggregate_{G}(\cdot)$ is a node aggregation function that is similar to the one for aggregating nodes' neighbors.

\begin{figure*}[t]
\begin{center}
\centerline{\includegraphics[width=1.7\columnwidth]{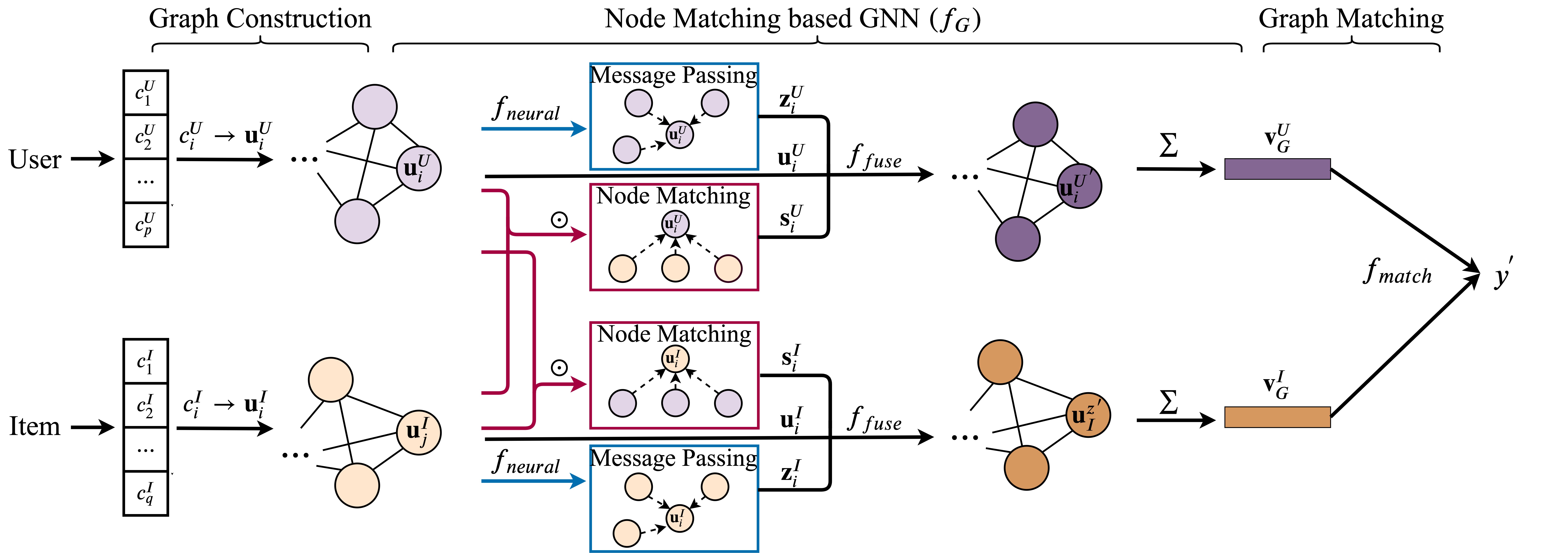}}
\vskip -0.1in
\caption{An Overview of the GMCF Model.}
\label{fig:gmcf_frame}
\end{center}
\vskip -0.10in
\end{figure*}

\section{Our Approach}
\label{sec:our_approach}

In this section, we describe our model in detail. First, we give an overview of our GMCF model. 
Then, we unveil the details of each part of the model.
Finally, we discuss the relations of our model to existing work and applicable situations.

\subsection{GMCF Overview}

	Our proposed model GMCF mainly consists of three modules: (i) User and Item Graph Construction Module,
	(ii) Node Matching based GNN Module, and (iii) Graph Representation Matching Module. 
	Figure \ref{fig:gmcf_frame} shows an overview of the GMCF model.
	In particular, for an input data sample $\bm{x} = \{c^U_i = (att, val)\}_{1 \leq i\leq p} \cup \{c^I_j = (att, val)\}_{1\leq j\leq q}$, each module works as follows.

	\vspace{1mm}
	\noindent
	\underline{The User and Item Graph Construction Module.}
	GMCF constructs a {\em user attribute graph} and an {\em item attribute graph} respectively based on the user and item characteristic specified in the data sample $\bm{x}$. 
	More specifically, the user attribute graph (resp., item attribute graph) is a {\em complete graph}, where each node corresponds to an attribute-value pair $(att, val)$ in the user (resp., item) characteristic and is represented by the representation $\bm{u} = val \cdot \bm{v}$ of the pair. 

	\vspace{1mm}
	\noindent
	\underline{The Node Matching based GNN Module.}
	For each node $i$ with representation $\bm{u}_i$ in the user (resp., item) attribute graph, this module first computes the {\em message passing information} $\bm{z}_i$ and the {\em node matching information} $\bm{s}_i$ with the item (resp., user) attribute graph.
	It then fuses $\bm{u}_i$, $\bm{z}_i$ and $\bm{s}_i$ to obtain a fused node representation $\bm{u}'_i$ for the node $i$.
	Finally, a graph representation $\bm{v}_G$ is obtained by aggregating the fused node representation $\bm{u}'_i$ for all nodes.

	\vspace{1mm}
	\noindent
	\underline{The Graph Representation Matching Module.}
	Our GMCF  performs a graph matching between the user and item attribute graph representations.
	The final prediction is obtained from matching result. 

Next, we introduce the details of the three modules.

\subsection{User \& Item Graph Construction}
We represent each user and each item as an \textit{attribute graph}, with their attributes as nodes and their inner interactions as edges. Specifically, for each data sample $\bm{x}$, the user attributes are the nodes of the user attribute graph.
Each node $i$ that represents an attribute takes the attribute representation $\bm{u}^U_i$ as the node representation.
Therefore, we represent the node set of the user attribute graph as $V^{U}=\{\bm{u}^U_i\}_{i\in \Ju}$. Then, each pair of nodes in the graph are connected with an edge to indicate the pairwise interaction between the two attributes. In summary, each user attribute graph is represented as $G^{U} = \la V^{U}, E^U \ra$, where $E^U$ is the edge set that contains all edges in the graph.
Note that since we consider all pairwise attribute interactions, the user attribute graph is a complete graph. 
We perform the same transformation for item attributes and get the item graph $G^{I} = \la V^{I}, E^I\ra$. 
In practice, we perform the graph construction by simply dividing attribute interactions as edges and matching pairs for different modules, which does not take additional effort than other explicit pairwise interaction modeling methods (e.g., AutoInt, Fi-GNN).

\subsection{Node Matching based GNN}

We propose a node matching based GNN, $f_G$ that considers both inner interactions through message passing and cross interactions through node matching. In this section, we describe how $f_G$ outputs the graph representation through modeling the two types of interactions. Note that since the modeling on user attribute graphs and item attribute graphs are symmetric, the notations for $f_{G}$ in the following subsections is generic (i.e., we omit the superscript $U$ and $I$) that can apply to both user and item attribute graphs.

\subsubsection{Neural Interaction based Message Passing}
In our model, the message passing method models the inner interactions for characteristic learning. Inspired by \cite{battaglia2016interaction, su2020detecting}, we use an MLP to model each inner interactions. Then, we aggregate the interaction modeling results as the message passing information. Specifically, an MLP function $f_{neural}\in\mathbb{R}^{2\times d}\rightarrow\mathbb{R}^d$, takes the embeddings of the two nodes as input and output the interaction modeling results:
\begin{equation}
\small
\label{fun:message_passing}
    \bm{z}_{ij} = f_{neural}(\bm{u}_i, \bm{u}_j),
\end{equation}
where $\bm{z}_{ij}$ is the interaction modeling results of the node pair ($i$, $j$).

Then, all the interaction modeling results corresponding to each node will be aggregated as the message passing information. In GMCF, we use the element-wise sum to aggregate the interaction modeling results $\bm{z}_{i} = \sum_{j\in N_i}\bm{z}_{ij}$, where $\bm{z}_{i}\in\mathbb{R}^d$ is the message passing results of node $i$ and $N_i$ is a set of node $i$'s neighbors.

$f_{neural}$ explicitly models the interaction information between two attributes, which effectively unifies the interaction modeling for recommendation and the message passing in graph learning.
In addition, inner interactions are used for capturing user (item) characteristics and are inherently complex. A high inner interaction result does not mean that the two attributes should be similar when determining the characteristics of a user (an item), which are different from cross interaction model results that reveal similarity (will be discussed in the next section). Therefore, a neural method (e.g., MLP) that non-linearly models the two attributes are desired.

\subsubsection{Bi-Interaction based Node Matching}

GMCF conducts node matching between two graphs through modeling cross interactions. Intuitively, we expect that an attribute $c^U_i$ will have a high matching score with an attribute $c^I_j$ if $c^U_i$ shows a high preference on $c^I_j$. For example, if male users prefer sci-fi movies, the node matching score of the node pairs $<$\textit{male}, \textit{sci-fi}$>$ should be high. In collaborative filtering, if a user attribute has a high preference for an item attribute, their embeddings should be similar after training.
To achieve this, we use the Bi-interaction \cite{he2017neural} for node matching, which keeps the monotonically increasing correlation between interaction modeling results and the attribute similarities.
As a result, if a user attribute has a high matching score on an item attribute, they have similar attribute representations.
Specifically, the Bi-interaction algorithm models the attribute interactions as:
\begin{equation}
\small
\label{fun:node_matching}
    \bm{s}_{ij} = \bm{u}_i \odot \bm{\hat{u}}_j,
\end{equation}
where $\bm{u}_i$ is the embedding of node $i$ in one graph, $\bm{\hat{u}}_j$ is the embedding of node $j$ in the other graph, $\bm{s}_{ij}$ is the node matching result of two nodes from different graphs, and $\odot$ is the element-wise product.

Similar to the message passing, we use an element-wise sum to aggregate the node matching result between a node in one graph and all nodes in the other graph. The aggregated node matching result is $\bm{s}_{i} = \sum_{j\in \hat{V}}\bm{s}_{ij}$, where 
$\hat{V}$ is the node set of the other attribute graph and
$\bm{s}_{i}$ is the aggregated node matching result of node $i$.

\subsubsection{Information Fusing}

Besides message passing results, GMCF further considers the node matching results to capture the node-level matching information while generating the fused node representations. Specifically, the node fusing function $f_{fuse}\in\mathbb{R}^{3\times d}\rightarrow\mathbb{R}^{d}$ takes the initial node representation $\bm{u}_i$, the message passing results $\bm{z}_{i}$ and the node matching results $\bm{s}_{i}$ as input, and get the fused node representation. Formally, we have
$\bm{u}^{'}_{i} = f_{fuse}(\bm{u}_i, \bm{z}_i, \bm{s}_i)$,
where $\bm{u}^{'}_{i}$ is the fused node representation of node $i$.

We can use any method for $f_{fuse}$, e.g., element-wise addition or recurrent neural network. Through testing, we find that the recurrent neural networks perform the best. We use GRU, an effective recurrent neural network model, as the function $f_{fuse}$. Therefore, $f_{fuse}$ considers $[\bm{u}_i, \bm{z}_i, \bm{s}_i]$ as the input sequence and the final output hidden layer of GRU is the fused node representation.

\subsubsection{Node Representation Aggregation}

The fused node representations of each graph are aggregated as the graph representation. We use the element-wise sum to aggregate the node representations. In summary, the $f_{G}$ in GMCF is:
\begin{equation}
\small
\label{fun:f_gnn}
    f_{G}(G, \hat{V}) = \sum_{i\in V}\bm{u}^{'}_{i},
\end{equation}

\subsection{Graph Matching}

While performing graph matching, we get the vector representations of the user attribute graph and the item attribute graph through $f_{G}$, respectively:
\begin{equation}
\small
\label{fun:graph_representation}
\begin{split}
    \bm{v}^{U}_{G} = f_{G}(G^{U}, V^{I}), \quad \bm{v}^{I}_{G} = f_{G}(G^{I}, V^{U}).
\end{split}
\end{equation}

We match the two graphs by using dot product as $f_{match}$ on the two graph representations to get the predicted output $y^{'}={\bm{v}^{U}_{G}}^{\top}\bm{v}^{I}_{G}$.

\subsection{Model Training}
\label{sec:learning}

While training, we use the $L_2$ norm to regularize all the parameters of GMCF. Therefore, the empirical risk minimization function of GMCF minimizes a loss function and an $L_2$ norm:
\begin{equation}
\small
\label{fun:ermf}
\begin{split}
    \mathcal{R}(\bm{\theta})=\frac{1}{N}\sum_{n=1}^{N}&\mathcal{L}(F_{GMCF}(\bm{x}_n; \bm{\theta}),y_{n}) + \lambda (\norm{\bm{\theta}}_2), \\
    &\bm{\theta}^{*} =\argmin_{\bm{\theta}} {\mathcal{R}(\bm{\theta})}, 
\end{split}
\end{equation}
where $F_{GMCF}$ is the prediction function of GMCF that outputs $y^{'}$, $\mathcal{L}(\cdot)$ corresponds to a loss function (e.g., binary cross-entropy
loss), $\bm{\theta}$ are all parameters in GMCF, and $\bm{\theta}^{*}$ are the final parameters.

\subsection{Relation to Existing Work and Discussion}

Our model has close relations to attribute graph-based recommender systems and FM. In this section, we discuss the relations to the two types of recommender systems. Then, we show that our model can apply to situations when the user or item attributes are not available, which happens in practice. 

\subsubsection{Relation to Attribute Graph-based Recommender Systems}
Fi-GNN \cite{li2019fi} and $L_0$-SIGN \cite{su2020detecting} are two attribute graph-based models that treat all attribute interactions equally in one graph. Our model can be considered an extension of this framework. Specifically, if GMCF considers all interactions equally, e.g., all are modeled as cross interactions, and puts the match function $f_{match}$ linear, e.g., $f_{match}(\bm{v}^{U}_{G}, \bm{v}^{I}_{G})=sum(\bm{v}^{U}_{G}) + sum(\bm{v}^{I}_{G})$, where $sum(\cdot)$ sums up all the elements of the input vector.
As a result, GMCF has a similar framework to Fi-GNN and $L_0$-SIGN.

\subsubsection{Relation to FM}
Different from graph-based models, FM aggregates all interaction modeling results linearly (i.e., sum up). GMCF can also be regarded as an extension of FM. Besides the modifications that reduce to the framework of Fi-GNN and $L_0$-SIGN (every interaction is modeled by element-wise product), if the fusing function $f_{fuse}$ is linear, e.g., $f_{fuse}(\bm{u}_i, \bm{z}_{i}, \bm{s}_{i})=\bm{u}_i + \frac{1}{2}\sum_{j\in V^{a}/\{i\}}\bm{s}_{ij}$, where $V^{a}=V^U \cup V^I$, $\bm{s}_{ij}=\bm{u}_i\odot\bm{u}_j$, the prediction function becomes:
\begin{equation}
\label{fun:GMCF_to_FM}
\small
\begin{split}
   y^{'} &= sum(\sum_{i \in V^U}(\bm{u}_i+ \frac{1}{2}\sum_{j\in V^{a}/\{i\}}\bm{s}_{ij})) + sum(\sum_{i \in V^I}(\bm{u}_i + \frac{1}{2}\sum_{j\in V^{a}/\{i\}}\bm{s}_{ij}))  \\
   &=\sum_{i \in V^{a}}sum(\bm{v}_i)\cdot val_i + \sum_{i\in V^{a}}\sum_{j\in V^{a}, j>i} \bm{v}_{i}^{\top}\bm{v}_{j} \cdot val_i\cdot val_j. 
\end{split}
\end{equation}

Equation \ref{fun:GMCF_to_FM} is similar to the prediction function of FM in its original paper (Equation 1 in \cite{rendle2010factorization}) with the weight parameter $\omega_i$ of each attribute $i$ being $sum(\bm{v}_i)$ and the bias term $\omega_0$ being omitted (Note that $sum(\bm{u}_i\odot\bm{u}_j)=\bm{v}_{i}^{\top}\bm{v}_{j}\cdot val_i\cdot val_j$). 

\subsubsection{When the User or Item Attributes are Unavailable}

In some situations, the user or item attributes are not available. GMCF is also applicable in these situations. Specifically, if one type of attribute is unavailable, e.g., user attributes, the user attribute graph becomes a single node (user ID). 
Then, no inner interaction modeling is in the user attribute graph,
i.e., $\bm{v}^U_{G}=\bm{u}_u^{'}=f_{fuse}(\bm{u}_u, \bm{s}_u)$, where $\bm{u}_u$ is the node representation of user ID. 
If both user and item attributes are unavailable, both the two attribute graphs are represented by a single node. Then, there is no inner interaction modeling but only the cross interaction modeling between user ID and item ID. We will evaluate our model in these scenarios in Section \ref{exp:attribute_unavaialble}.

\section{Experiments}

In this section, we conduct experiments to evaluate GMCF. We focus on three questions: 
(i) what is the ability of GMCF in providing accurate recommendations comparing to other baselines that can consider user attributes and item attributes; 
(ii) what is the effectiveness of each component in GMCF on providing accurate predictions;
(iii) whether the learned attribute embeddings reveal collaborative relations and semantic meaning in their latent space.

\subsection{Experimental Protocol}

We first describe the datasets and the baselines used in our experiments. Then, we illustrate the experimental set-up in detail.

\subsubsection{Datasets}

We run GMCF and baselines on three datasets, that contain both user and item attributes. 
Table \ref{tab:dataset} shows their statistic information.
Below are the descriptions of the datasets:

\textbf{MovieLens 1M} \cite{harper2015movielens} contains users' ratings on movies. 
Each data sample contains a user and a movie with their corresponding attributes.
We further collect movies' other attributes, such as directors and casts from IMDB
to enrich the datasets.
\textbf{Book-crossing} \cite{ziegler2005improving} contains users' implicit and explicit ratings of books. Each data sample contains a user and a book with their corresponding attributes. The reprocessed words in the book titles are also regarded as attributes of the book.
\textbf{Taobao} \cite{zhou2018deep} is a dataset that collects the log of click on display advertisement displayed on the website of Taobao. Each log contains a user with corresponding attributes such as gender and age level, a displayed advertisement with attributes such as category and brand of the advertised item.

MovieLens 1M and Book-crossing contain explicit ratings. We transfer the explicit ratings to implicit feedback. We regard the ratings greater than 3 as positive ratings for MovieLens 1M and regard all rated explicit ratings as positive ratings for Book-crossing due to its sparsity. Then, we randomly select the same number of negative samples equal to the number of positive samples for each user. To ensure the datasets' quality, we select the users with more than 10 positive ratings for MovieLens 1M and have more than 20 positive ratings for Book-crossing and Taobao. 

\begin{table}[t]
\caption{Dataset statistics. The \textit{attr.} refers to "attributes".}
\small
\label{tab:dataset}
\begin{center}
\vskip -0.1in
\begin{tabular}{P{1.8cm}cccP{0.6cm}P{0.6cm}}
\hline
\multirow{2}{*}{\textbf{Dataset}} & \multirow{2}{*}{\#\textbf{Data}} & \multirow{2}{*}{\#\textbf{User}} & \multirow{2}{*}{\#\textbf{Item}} & \#\textbf{User} & \#\textbf{Item} \\
 & & & & \textit{attr.} & \textit{attr.} \\
\hline
MovieLens 1M & 1,149,238 & 5,950  & 3,514 & 30 & 6,944 \\
Book-Crossing &  1,050,834 & 4,873 & 53,168 & 87  & 43,157 \\
Taobao &  2,599,463 & 4,532 & 371,760 & 36  & 434,254 \\
\hline
\end{tabular}
\end{center}
\vskip -0.2in
\end{table}

\subsubsection{Baselines}

We compare our model with competitive baselines that can take user attributes and item attributes into account:

\textbf{FM} \cite{rendle2010factorization} models every feature interaction by dot product and sums up all the modeling results as the final prediction result.
\textbf{AFM} \cite{xiao2017attentional} additionally calculates an attention value for each interaction in FM as the weight when performing the aggregation.
\textbf{NFM} \cite{he2017neural} leverages an MLP on top of the aggregated interaction modeling results of FM so that as to non-linearly analyze the feature interactions.
\textbf{W\&D} \cite{cheng2016wide} combines a linear model that with a deep neural network for recommendation. The input of the deep neural network is the concatenation of all feature embeddings.
\textbf{DeepFM} \cite{guo2017deepfm} combines interaction analysis results from using MLP and FM for prediction. Specifically, in the MLP part, all attribute embeddings are concatenated together and then fed into an MLP to learn their interactions.
\textbf{AutoInt} \cite{song2019autoint} explicitly models all feature interactions using a multi-head self-attentive neural network and then aggregates all the modeling results as the final prediction result. 
\textbf{Fi-GNN} \cite{li2019fi} represents each data sample as a feature graph that each node is a feature field. Then, it models the interactions of the field in a GNN using the multi-head self-attention method.
\textbf{$L_0$-SIGN} \cite{su2020detecting} considers each data sample as a graph, with the user, the item, and all corresponding features as nodes. 
$L_0$-SIGN simultaneously detects beneficial interactions and leverages the detected ones as edges for graph classification, and the classification results are the prediction results. 
For all baselines, we set the MLP structure (if used) for interaction modeling to be the same as the MLP for inner interaction modeling in our model for a fair comparison.

\begin{table*}[ht]
\caption{Summary of the performance in comparison with baselines.}
\label{tab:performance}
\small
\begin{center}
\vskip -0.1in
\begin{tabular}{l|cccc|cccc|cccc}
\hline
 & \multicolumn{4}{c|}{\textbf{MovieLens 1M}} & \multicolumn{4}{c|}{\textbf{Book-Crossing}}  & \multicolumn{4}{c}{\textbf{Taobao}}\\
 & AUC & Logloss & NDCG@5 & NDCG@10 & AUC & Logloss & NDCG@5 & NDCG@10 & AUC & Logloss & NDCG@5 & NDCG@10\\
\hline
FM           & 0.8761 & 0.4409 & 0.8143 & 0.8431 & 0.7417 & 0.5771 & 0.7616 & 0.8029 & 0.6171 & 0.2375 & 0.0812 & 0.1120\\
AFM          & 0.8837 & 0.4323 & 0.8270 & 0.8676 & 0.7541 & 0.5686 & 0.7820 & 0.8258 & 0.6282 & 0.2205 & 0.0872 & 0.1240\\		
NFM          & 0.8985 & 0.3996 & 0.8486 & 0.8832 & 0.7988 & 0.5432 & 0.7989 & 0.8326 & \underline{0.6550} & 0.2122 & 0.0997 & 0.1251\\				
W\&D         & 0.9043 & 0.3878 & 0.8538 & 0.8869 & 0.8105 & 0.5366 & 0.8048 & 0.8381 & 0.6531 & 0.2124 & 0.0959 & 0.1242\\ 			
DeepFM       & 0.9049 & 0.3856 & 0.8510 & 0.8848 & 0.8127 & 0.5379 & 0.8088 & 0.8400 & \underline{0.6550} & \underline{0.2115} & 0.0974 & 0.1243\\
AutoInt      & 0.9034 & 0.3883 & 0.8619 & 0.8931 & 0.8130 & 0.5355 & 0.8127 & 0.8472 & 0.6434 & 0.2146 & 0.0924 & 0.1206\\			
\hline
Fi-GNN       & 0.9063 & 0.3871 & 0.8705 & 0.9029 & 0.8136 & 0.5338 & 0.8094 & 0.8522 & 0.6462 & 0.2131 & 0.0986 & 0.1241\\
$L_0$-SIGN   & \underline{0.9072} & \underline{0.3846} & \underline{0.8849} & \underline{0.9094} & \underline{0.8163} & \underline{0.5274} & \underline{0.8148} & \underline{0.8629} & 0.6547 & 0.2124 & \underline{0.1006} & \underline{0.1259}\\
\hline
GMCF & \textbf{0.9127} & \textbf{0.3789} & \textbf{0.9374} & \textbf{0.9436} & \textbf{0.8228} & \textbf{0.5233} & \textbf{0.8671} & \textbf{0.8951} & \textbf{0.6679} & \textbf{0.1960} & \textbf{0.1112} & \textbf{0.1467}\\ 		
\hline
\textit{Improv}  &  0.61\% & 1.48\% & 5.91\% & 4.28\% & 0.80\% & 0.78\% & 6.42\% & 3.73\% & 1.96\% & 7.32\% & 10.53\% & 16.52\%\\
\textit{p-value}  &  1.09\% & 0.25\% & 0.25\% & 0.25\% & 0.46\% & 0.25\% & 0.25\% & 0.25\% & 0.25\% & 0.25\%& 0.25\%& 0.25\%\\       
\hline
\end{tabular}
\end{center}
\vskip -0.10in
\end{table*}

\subsubsection{Experimental Set-Up}
We randomly split each dataset into training, validation, and test set for each user with a ratio of 6:2:2. The validation set is only used to decide the best parameter setting, and the test set is only used to evaluate the models.
Unless otherwise specified, we use the following hyper-parameter settings: the node representation dimension is 64 (i.e., $d=64$); the number of hidden layers for MLP is 1 and the number of units for the hidden layer is $4d$; the learning rate is $1\times 10^{-3}$; the $\lambda$ for the regularization is $1\times 10^{-5}$. 
We use the binary cross entropy as the loss function and use Adam \cite{kingma2014adam} as the optimization algorithm. 
We use the area under the curve (AUC), Logloss, and NDCG@k as evaluation metrics to evaluate the performance of our model and baseline models. AUC and Logloss are frequently used in the tasks with implicit feedback and NDCG@k is a frequently used metric to evaluate the top-k recommendation. We set k to 5 and 10. All the experiments are repeated 5 times and the average results are recorded.

\subsection{Overall Performance}

We compare the performance of GMCF with the baselines. Table \ref{tab:performance} shows the prediction performance of each model. The best results for each dataset are in bold, and the best baseline results are in underline. The rows \textit{Improv} and \textit{p-value} show the improvement and statistical significance test results (through Wilcoxon signed-rank test) of GMCF and the best baseline results, respectively.
From Table \ref{tab:performance}, we observe that:
\begin{itemize}[leftmargin = *]
    \item GMCF outperforms all the baselines significantly, with the p-value of all metrics rejecting the null hypothesis with a level of significance of $\alpha=5\%$.
    These results prove the ability of GMCF in effectively analyzing the structural information of user and item attributes for accurate predictions.
    \item The models that explicitly model attribute interactions (GMCF, AutoInt, Fi-GNN, $L_0$-SIGN) gain better prediction accuracy than other models. This indicates that the explicit interaction modeling is promising to extract useful information from attributes and attribute interactions for accurate predictions.
    \item FM and AFM perform worse than the other models. This is because FM and AFM do not use neural methods, but solely rely on the dot product to extract information from attribute interactions. Hence, to model more complicated interactions, sophisticated methods (e.g., MLP) are required. Accordingly, our model leverages MLP to model the inner interactions inside the user attribute graph and the item attribute graph, which provides powerful interaction modeling ability for accurate predictions.
    \item The GNN-based models (GMCF, Fi-GNN, $L_0$-SIGN) gain better prediction accuracies than other models. It shows the ability of GNNs in modeling attribute interactions for recommendation. GMCF further models attribute interactions in a structure of graph matching, which is more suitable for recommendation and gains better performance than Fi-GNN and $L_0$-SIGN.
\end{itemize}

\subsection{Study of Neural Graph Matching}

In this section, We evaluate the neural graph matching of GMCF. We focus on (i) the effectiveness of modeling inner and cross interactions using different methods; (ii) the comparison of using different algorithms as the fusing function; (iii) the performance of GMCF when the user and item attributes are not available. Due to space limitation, we omit the results of Logloss and NDCG@5, which show similar trends with AUC and NDCG@10, respectively.

\subsubsection{Evaluation of Inner and Cross Interaction modeling}
\label{sec:eval_inner_cross}

We evaluate the inner and cross interaction modeling methods. Specifically, we focus on three questions: 1) what is the effectiveness of node-level graph matching (i.e., modeling cross interactions)?
2) whether we should use different methods to model the two types of interactions?
3) whether sophisticated nonlinear algorithms are always suitable for modeling the two types of interactions?

To answer the three questions, we run GMCF with different interaction modeling methods. For clear demonstration, we use the pair <\textit{inner interaction model, cross interaction model}> to indicate the method combinations in each variation. For example, we represent the original GMCF as <\textit{MLP, Bi}> to indicate an MLP-based message passing and a Bi-interaction-based node matching.

\begin{figure}
\centering
\begin{tikzpicture}
\begin{axis}[xbarstyle, title=MovieLens 1M, xlabel={AUC}]
\addplot [color=mycolor1, fill=mycolor1] coordinates { 
(0.8742,<\textit{MLP, None}>)
(0.9127,\textbf{<\textit{MLP, Bi}>})         
(0.9025,<\textit{Bi, Bi}>)              
(0.8964,<\textit{MLP, MLP}>)            
(0.9054,<\textit{MLP1, MLP2}>)          
(0.8941,<\textit{Bi, MLP}>)             
};
\end{axis}
\end{tikzpicture}
\begin{tikzpicture}
\begin{axis}[xbarstyle,title=MovieLens 1M, xlabel={NDCG@10}]
\addplot[color=mycolor1, fill=mycolor1] coordinates { 
(0.9072,<\textit{MLP, None}>)
(0.9436,\textbf{<\textit{MLP, Bi}>})   
(0.9342,<\textit{Bi, Bi}>)             
(0.9257,<\textit{MLP, MLP}>)           
(0.9391,<\textit{MLP1, MLP2}>)         
(0.9201,<\textit{Bi, MLP}>)            
};
\end{axis}
\end{tikzpicture}
\begin{tikzpicture}
\begin{axis}[xbarstyle, title=Taobao, xlabel={AUC}]
\addplot[color=mycolor1, fill=mycolor1] coordinates {
(0.6501,<\textit{MLP, None}>)               
(0.6679,\textbf{<\textit{MLP, Bi}>})         
(0.6608,<\textit{Bi, Bi}>)                  
(0.6639,<\textit{MLP, MLP}>)                
(0.6647,<\textit{MLP1, MLP2}>)               
(0.6563,<\textit{Bi, MLP}>)                 
};
\end{axis}
\end{tikzpicture}
\begin{tikzpicture}
\begin{axis}[xbarstyle, title=Taobao, xlabel={NDCG@10}]
\addplot[color=mycolor1, fill=mycolor1] coordinates {
(0.6708,<\textit{MLP, None}>)            
(0.7059 ,\textbf{<\textit{MLP, Bi}>})         
(0.6852,<\textit{Bi, Bi}>)                
(0.6891,<\textit{MLP, MLP}>)             
(0.6936,<\textit{MLP1, MLP2}>)           
(0.6850,<\textit{Bi, MLP}>)              
};
\end{axis}
\end{tikzpicture}
\vskip -0.1in
\caption{The comparison of using different inner interaction and cross interaction modeling combinations.}
\label{fig:interaction_eva}
\vskip -0.1in
\end{figure}
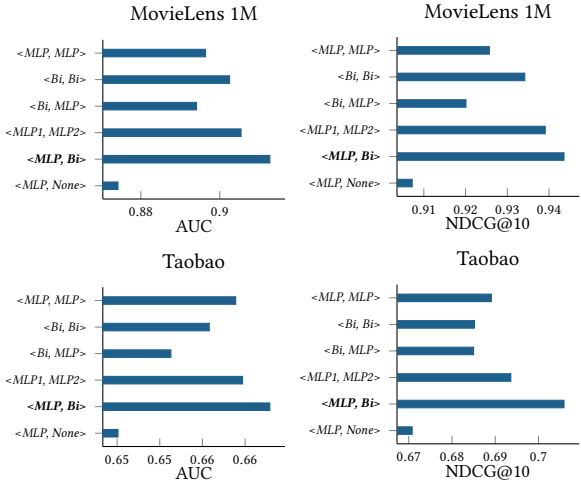

We run GMCF using the combination <\textit{MLP, None}>, where \textit{None} indicates that we do not perform node-level matching (i.e., no cross interaction modeling). Then, we run two variations that use the same method to model all interactions: <\textit{Bi, Bi}>, <\textit{MLP, MLP}>. Note that the second variation uses the same MLP (i.e., the same neural architecture with shared parameters). Finally, we further run two variations that use different methods: <\textit{MLP1, MLP2}> (the same architecture with different parameters) and <\textit{Bi, MLP}>. 
Figure \ref{fig:interaction_eva} shows the results of using different variations. We omit the results of Book-crossing due to the space limitation, which shows similar trend with MovieLens 1M (the same as the remaining figures). 

For question 1), <\textit{MLP, None}> gains the worst performance. The reason is that the attribute graphs are inherently complex. Fusing this information only in the last step makes it difficult to match users' preferences on items through attributes. The node matching method provides the explicit attribute communication through the cross interactions between the two graphs, which results in a fine-grained analysis for a more accurate preference matching.
For question 2), the combinations (<\textit{Bi, Bi}> and <\textit{MLP, MLP}>) gain relatively worse results than the original setting (<\textit{MLP, Bi}>).
This is because the inner interactions and cross interactions are inherently different that should be modeled differently that fit their roles. The inner interactions are used for characteristic learning and do not indicate the similarity information (i.e., two attributes having strong interaction does not mean that they are similar). Meanwhile, in collaborative filtering, two attributes are expected to show similarity information if their cross interaction is strong (high preference). The Bi-interaction algorithm fits for the requirement of cross interactions.
For question 3), even using different methods, <\textit{MLP1, MLP2}> and <\textit{Bi, MLP}> still gain worse results than the original GMCF. Although an MLP seems more powerful than the Bi-interaction algorithm, MLP cannot to effectively learn the measurement for the cross interactions. The results show that sophisticated algorithms such as MLP are expected to model the inner interactions, and similarity measurements such as the Bi-interaction algorithm are expected to model the cross interactions. 

\vskip -0.1in
\subsubsection{Evaluation of the Fusing Method}

The fusing method in GMCF aggregates the message passing results and the node matching results to get the fused node representation.
We evaluate the effectiveness of using different algorithms as the fusing algorithm. Except the GRU, we further use the element-wise sum on the three vectors as the fused node representation (SUM) and the MLP that concatenates the vectors as input and outputs the fused node representation (MLP). We use the MLP that has one hidden layer, with the number of neurons being $4d$. 

Table \ref{tab:node_update} shows the experimental results of using the three algorithms as the fusing method. From the table, we can see that using GRU results in the best performance on all datasets. It shows the ability of GRU to effectively aggregate the message passing information and the node matching information into the fused node representation for accurate predictions. Summing up the results (\textit{SUM}) gains the worst results in most situations, which indicates that the message passing and node matching information are complex. Powerful algorithms are required to fuse them.

\begin{table}[t]
\caption{The performance of using different fusing algorithms.}
\label{tab:node_update}
\footnotesize
\begin{center}
\vskip -0.13in
\begin{tabular}{l|cc|cc|cc}
\hline
 & \multicolumn{2}{c|}{\textbf{MovieLens 1M}} & \multicolumn{2}{c|}{\textbf{Book-Crossing}} & \multicolumn{2}{c}{\textbf{Taobao}}\\
 & AUC & NDCG@10 & AUC & NDCG@10 & AUC & NDCG@10\\
\hline
SUM & 0.9042 &	0.9404 & 0.8190 & 0.8904 & 0.6583 &	0.1383\\
MLP & 0.9022 & 0.9389 & 0.8156	& 0.8920 & 0.6636 &	0.1421\\
GRU & \textbf{0.9127} & \textbf{0.9436} & \textbf{0.8228} & \textbf{0.8951} & \textbf{0.6679} & \textbf{0.1467}\\
\hline
\end{tabular}
\end{center}
\vskip -0.1in
\end{table}

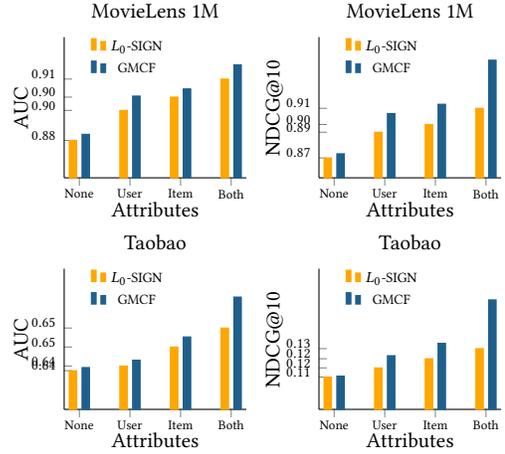
\begin{figure}[t]
\centering
\begin{tikzpicture}
\begin{axis}[barstyle, title=MovieLens 1M, xlabel={Attributes}, ylabel={AUC}, legend style={draw=none, at={(0.1,0.85)},anchor=west, nodes={scale=0.9, transform shape}, legend image post style={scale=0.6}},ymin=0.87, ymax=0.923]
\addplot [color=mycolor4, fill=mycolor4] coordinates { 
(None, 0.8841)         
(User, 0.8954)
(Item, 0.9004)  
(Both, 0.9072)
};
\addplot[color=mycolor1, fill=mycolor1] coordinates { 
(None, 0.8864)         
(User, 0.9008)
(Item, 0.9035)  
(Both, 0.9126) 
};
\legend{$L_0$-SIGN,GMCF}
\end{axis}
\end{tikzpicture}
\begin{tikzpicture}
\begin{axis}[barstyle, title=MovieLens 1M, xlabel={Attributes}, ylabel={NDCG@10}, legend style={draw=none, at={(0.1,0.85)},anchor=west, nodes={scale=0.9, transform shape}, legend image post style={scale=0.6}}, ymin=0.86, ymax=0.96]
\addplot [color=mycolor4, fill=mycolor4] coordinates { 
(None, 0.8741)         
(User, 0.8924)
(Item, 0.8979)  
(Both, 0.9094)
};
\addplot[color=mycolor1, fill=mycolor1] coordinates { 
(None, 0.8771)         
(User, 0.9057)
(Item, 0.9123)  
(Both, 0.9436)  
};
\legend{$L_0$-SIGN,GMCF}
\end{axis}
\end{tikzpicture}
\begin{tikzpicture}
\begin{axis}[barstyle, title=Taobao, xlabel={Attributes}, ylabel={AUC}, legend style={draw=none, at={(0.1,0.85)},anchor=west, nodes={scale=0.9, transform shape}, legend image post style={scale=0.6}}, ymin=0.62, ymax=0.68]
\addplot [color=mycolor4, fill=mycolor4] coordinates { 
(None, 0.6365)         
(User, 0.6385)
(Item, 0.6466)  
(Both, 0.6547)
};
\addplot[color=mycolor1, fill=mycolor1] coordinates { 
(None, 0.63784)         
(User, 0.6410)
(Item, 0.6509)  
(Both, 0.6679) 
};
\legend{$L_0$-SIGN,GMCF}
\end{axis}
\end{tikzpicture}
\begin{tikzpicture}
\begin{axis}[barstyle, title=Taobao, xlabel={Attributes},ylabel={NDCG@10}, legend style={draw=none, at={(0.1,0.85)},anchor=west, nodes={scale=0.9, transform shape}, legend image post style={scale=0.6}}, ymin=0.1, ymax=0.16]
\addplot[color=mycolor4, fill=mycolor4] coordinates { 
(None, 0.1138)         
(User, 0.1177)
(Item, 0.1216)  
(Both, 0.1259) 
};
\addplot[color=mycolor1, fill=mycolor1] coordinates { 
(None, 0.1142)         
(User, 0.1229)
(Item, 0.1282)  
(Both, 0.1467) 
};
\legend{$L_0$-SIGN,GMCF}
\end{axis}
\end{tikzpicture}
\vskip -0.10in
\caption{The comparison when using different attributes.}
\label{fig:user_item_not_aval}
\vskip -0.15in
\end{figure}

\subsubsection{Evaluation of the User and Item Attributes}
\label{exp:attribute_unavaialble}

In some situations, user or item attributes are not available. We evaluate how GMCF performs in these situations. 
GMCF is a flexible framework that is applicable when the user or item attributes are not available. In these situations, the user (or item) graphs are reduced into a single node indicating user (or the item) ID.
We run GMCF and the best baseline $L_0$-SIGN on the situations that user or item attributes are not available. Figure \ref{fig:user_item_not_aval} shows the results. Specifically, in the x-axis, ``\textit{None}'' indicates neither user or item attributes are available, ``\textit{User}'' indicates only user attributes are available, ``\textit{Item}'' indicates only item attributes are available, and ``\textit{Both}'' indicates both user and item attributes are available.

From the figure, we observe that: 
1) Both models perform better when user and item attributes become available (from left to right). It shows that both user and item attributes are useful for performance gain. Although the performance gain from user attributes (\textit{User}) seems not that significant compared to item attributes (\textit{Item}), the user attributes provide useful information for potential explanations of the prediction results, which will be discussed in Section \ref{subsec:case_study}.
2) The two models have similar performance when no attributes are available (\textit{None}). However, GMCF performs much better when user and item attributes are available. The performance gain compared to $L_0$-SIGN when the user and item attributes are available resulted from the generated graph matching structure, which captures more useful structural information for accurate predictions. 

\begin{figure*}[ht]
\centering
\begin{subfigure}[c]{0.255\textwidth}
\centerline{\includegraphics[width=1\columnwidth]{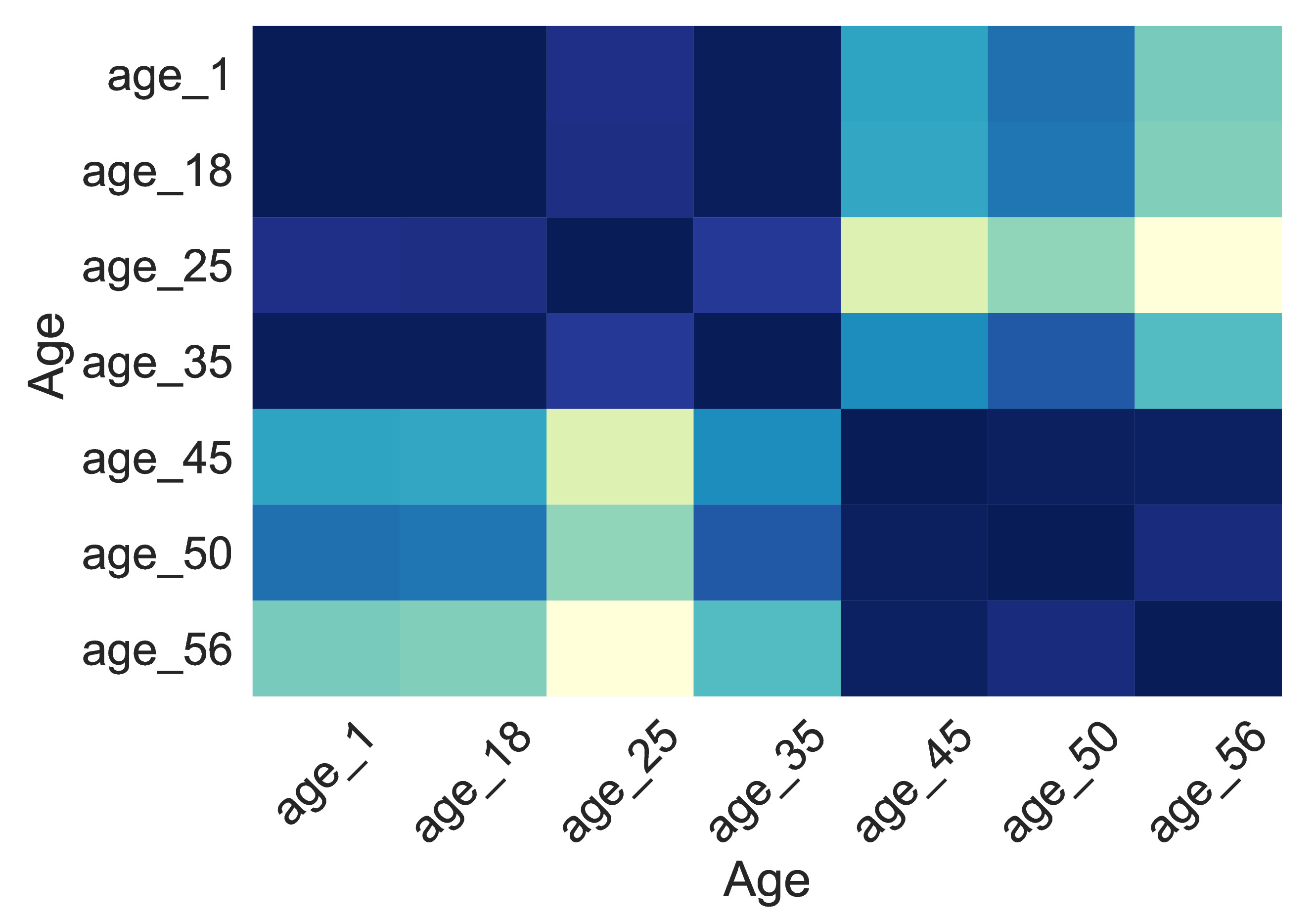}}
\end{subfigure}
\label{fig:interpretation}
\begin{subfigure}[c]{0.255\textwidth}
\centerline{\includegraphics[width=1\columnwidth]{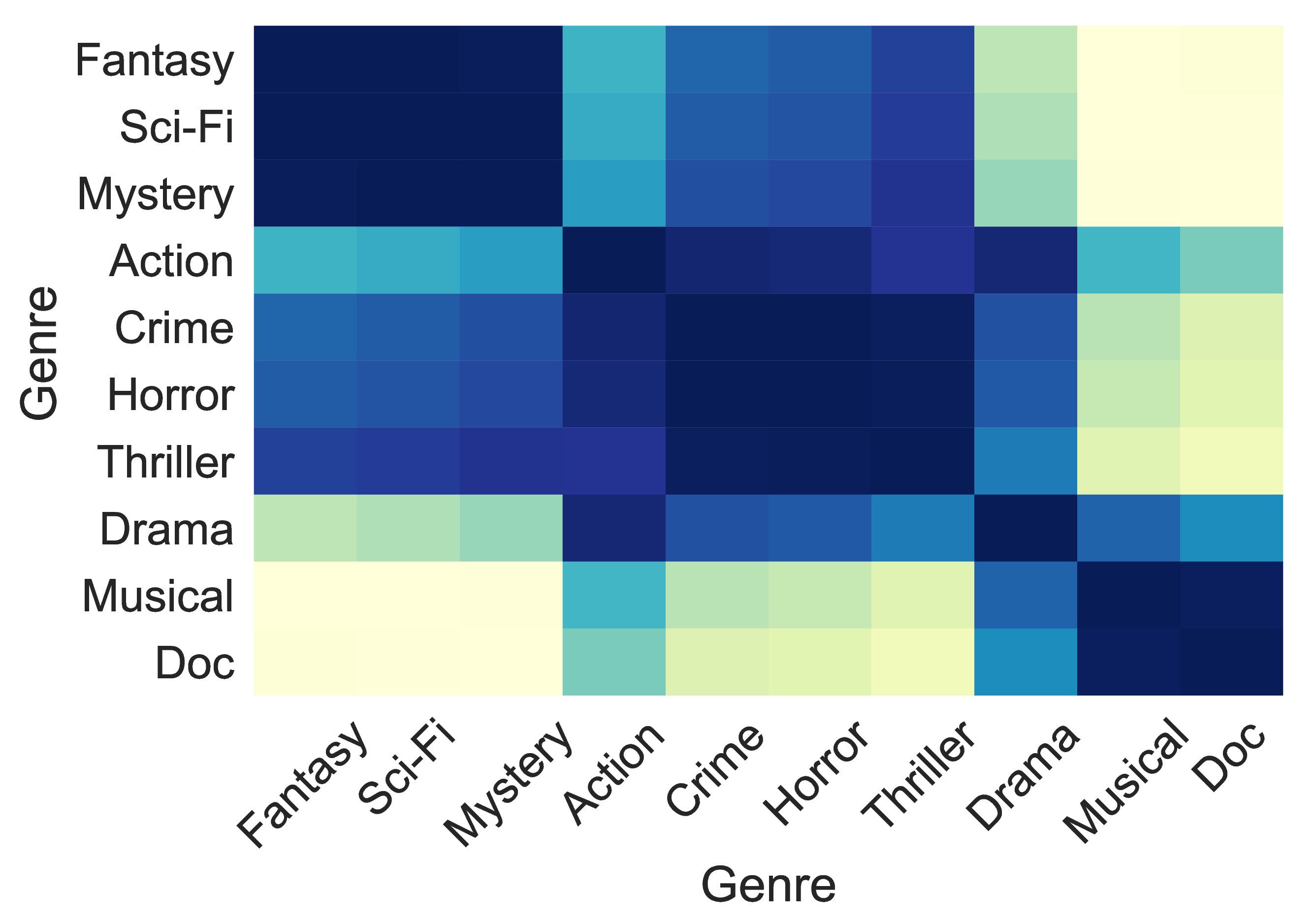}}
\end{subfigure}
\begin{subfigure}[c]{0.255\textwidth}
\centerline{\includegraphics[width=1\columnwidth]{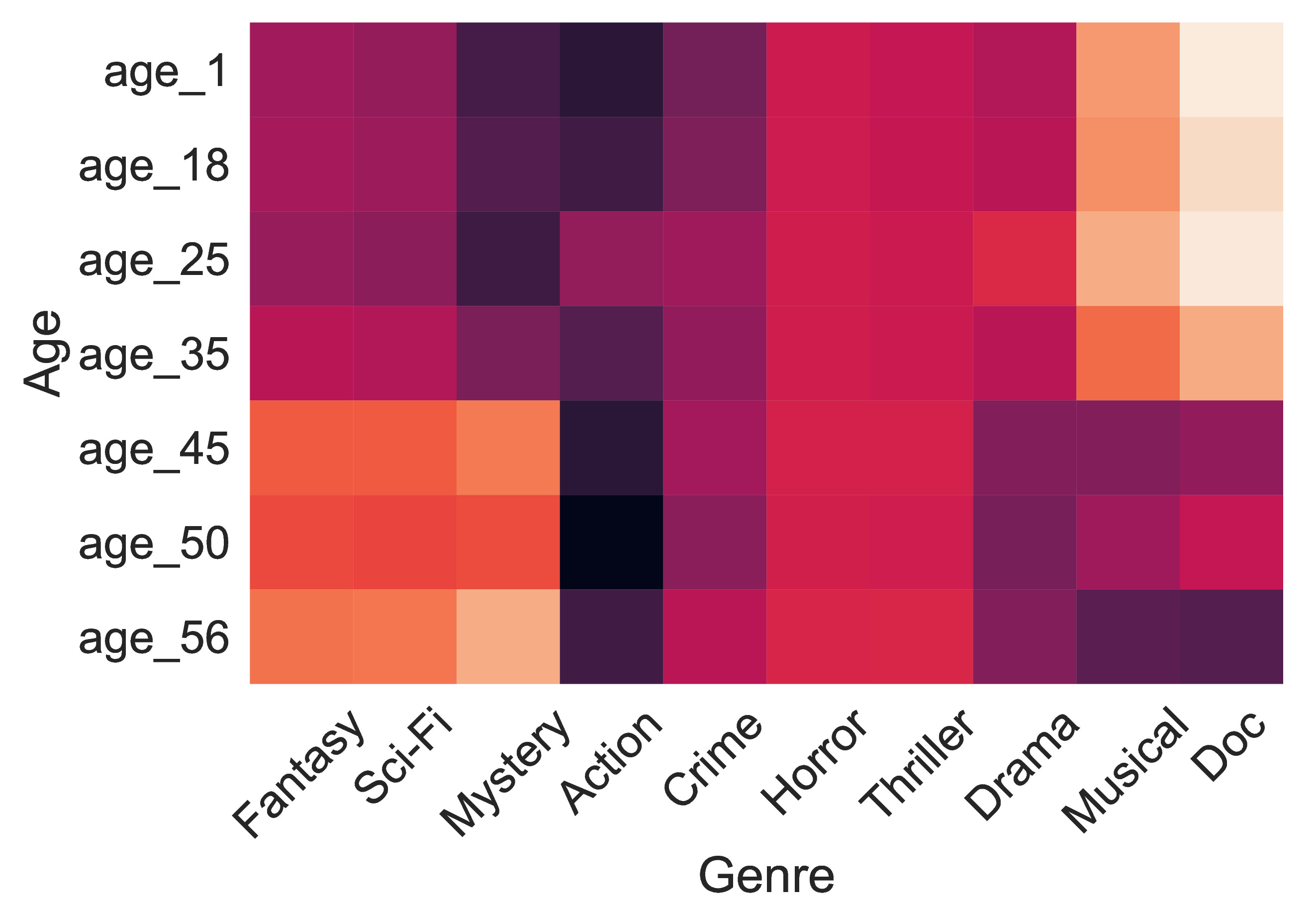}}
\end{subfigure}
\vskip -0.15in
\caption{The case studies on the MovieLens 1M dataset. \textit{Left}: the embedding similarities between user age groups. \textit{Central}: the embedding similarities between movie genres. \textit{Right}: the node matching (preference) results between the ages groups and the genres.}
\vskip -0.13in
\label{fig:case_study}
\end{figure*}

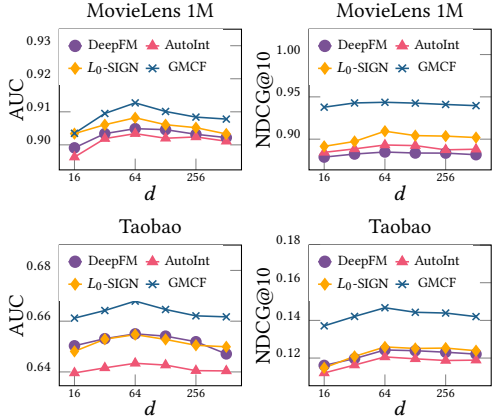
\begin{figure}[ht]
\begin{center}
\begin{tikzpicture}
\begin{axis}[mystyle2, title =MovieLens 1M, ylabel={AUC}, xlabel={$d$},legend columns=2, ymax=0.935, symbolic x coords={16,32,64,128,256, 512}, legend style={draw=none, at={(0.9,0.80)},anchor=east, nodes={scale=0.9, transform shape}}, legend image post style={scale=0.6}]
\addplot[mark=*, color=mycolor2] coordinates {
	(16, 0.8991)
	(32, 0.9034)
	(64, 0.9049)
	(128, 0.9046)
	(256, 0.9032)
	(512, 0.9022)
};
\addplot[mark=triangle*,color=mycolor3] coordinates {
	(16, 0.8963)
	(32, 0.9019)
	(64, 0.9034)
	(128, 0.9020)
	(256, 0.9024)
	(512, 0.9011)
};

\addplot[mark=diamond*, color=mycolor4] coordinates {
	(16, 0.9034)
	(32, 0.9061)
	(64, 0.9082)
	(128, 0.9061)
	(256, 0.9052)
	(512, 0.9033)
};
\addplot[mark=x,color=mycolor1] coordinates {
	(16, 0.9035)
	(32, 0.9095)
	(64, 0.9127)
	(128, 0.9101)
	(256, 0.9084)
	(512, 0.9078)
};
\legend{DeepFM, AutoInt, $L_0$-SIGN, GMCF}
\end{axis}
\end{tikzpicture}
\begin{tikzpicture}
\begin{axis}[mystyle2, title =MovieLens 1M, ylabel={NDCG@10}, xlabel={$d$}, symbolic x coords={16,32,64,128,256,512}, ymax=1.03, legend columns=2,legend style={draw=none, at={(0.9,0.80)},anchor=east, nodes={scale=0.9, transform shape}}, legend image post style={scale=0.6}]
\addplot[mark=*, color=mycolor2] coordinates {
	(16, 0.8788)
	(32, 0.8824)
	(64, 0.8848)
	(128, 0.8836)
	(256, 0.8835)
	(512, 0.8816)
};
\addplot[mark=triangle*,color=mycolor3] coordinates {
	(16, 0.8844)
	(32, 0.8885)
	(64, 0.8931)
	(128, 0.8924)
	(256, 0.8874)
	(512, 0.8881)
};
\addplot[mark=diamond*, color=mycolor4] coordinates {
	(16, 0.8915)
	(32, 0.8973)
	(64, 0.9094)
	(128, 0.9042)
	(256, 0.9036)
	(512, 0.9019)
};
\addplot[mark=x,color=mycolor1] coordinates {
	(16, 0.9378)
	(32, 0.9428)
	(64, 0.9436)
	(128, 0.9426)
	(256, 0.9410)
	(512, 0.9396)
};
\legend{DeepFM, AutoInt, $L_0$-SIGN, GMCF}
\end{axis}
\end{tikzpicture}
\begin{tikzpicture}
\begin{axis}[mystyle2, title =Taobao, ylabel={AUC}, xlabel={$d$}, symbolic x coords={16,32,64,128,256, 512}, ymax=0.69, legend columns=2,legend style={draw=none, at={(0.9,0.80)},anchor=east, nodes={scale=0.9, transform shape}}, legend image post style={scale=0.6}]
\addplot[mark=*, color=mycolor2] coordinates {
	(16, 0.6503)
	(32, 0.6531)
	(64, 0.6550)
	(128, 0.6541)
	(256, 0.6519)
	(512, 0.6471)
};
\addplot[mark=triangle*,color=mycolor3] coordinates {
	(16, 0.6396)
	(32, 0.6416)
	(64, 0.6434)
	(128, 0.6427)
	(256, 0.6405)
	(512, 0.6404)
};
\addplot[mark=diamond*,color=mycolor4] coordinates {
	(16, 0.6481)
	(32, 0.6529)
	(64, 0.6547)
	(128, 0.6528)
	(256, 0.6505)
	(512, 0.6499)
};
\addplot[mark=x,color=mycolor1] coordinates {
	(16, 0.6612)
	(32, 0.6642)
	(64, 0.6679)
	(128, 0.6646)
	(256, 0.6621)
	(512, 0.6617)
};
\legend{DeepFM, AutoInt, $L_0$-SIGN, GMCF}
\end{axis}
\end{tikzpicture}
\begin{tikzpicture}
\begin{axis}[mystyle2, title =Taobao, ylabel={NDCG@10}, xlabel={$d$}, symbolic x coords={16,32,64,128,256, 512}, ymax=0.18, legend columns=2,legend style={draw=none, at={(0.9,0.80)},anchor=east, nodes={scale=0.9, transform shape}}, legend image post style={scale=0.6}]
\addplot[mark=*,color=mycolor2] coordinates {
    (16, 0.1161)
	(32, 0.1197)
	(64, 0.1243)
	(128, 0.1239)
	(256, 0.1231)
	(512, 0.1221)
};
\addplot[mark=triangle*,color=mycolor3] coordinates {
    (16, 0.1121)
	(32, 0.1163)
	(64, 0.1206)
	(128, 0.1196)
	(256, 0.1187)
	(512, 0.1189)
};
\addplot[mark=diamond*, color=mycolor4] coordinates {
	(16, 0.1148)
	(32, 0.1209)
	(64, 0.1259)
	(128, 0.1251)
	(256, 0.1253)
	(512, 0.1239)
};
\addplot[mark=x,color=mycolor1] coordinates {
    (16, 0.1371)
	(32, 0.1421)
	(64, 0.1467)
	(128, 0.1443)
	(256, 0.1439)
	(512, 0.1420)
};
\legend{DeepFM, AutoInt, $L_0$-SIGN, GMCF}
\end{axis}
\end{tikzpicture}
\vskip -0.1in
\caption{The performance of GMCF and selected baselines in different node (attribute) embedding dimensions.}
\label{fig:parameter_learning}
\end{center}
\vskip -0.1in
\end{figure}

\subsection{Parameter study}

In this section, we evaluate GMCF with different hyper-parameter settings. Specifically, we evaluate the node representations' dimension and the depth of the MLP used in our model.

Figure \ref{fig:parameter_learning} shows the results of our model and best-performed baselines on different node representation dimensions ($d$).
From the figure, we observe that GMCF constantly outperforms baselines on different node representation dimensions, which shows the robustness of our model in delivering superior prediction accuracy. Then, when the dimension is 64, our model and most of the baselines gain the best performance. This indicates that a highger dimension does not necessarily result in better prediction accuracy. This is because that a larger dimension means more parameters to fit, and thus is prone to cause the overfitting problem.

Then, we evaluate how the different number of hidden layers in the MLP affects our model's performance. In our model, we use an MLP with 1 hidden layer to analyze the inner interactions while message passing. Now we evaluate our model with different number of layers. Specifically, we run our model with $0,1,2,3,4$ hidden layers respectively (GMCF-0,...,GMCF-4). Note that GMCF-0 means that the MLP only performs a linear transformation from the node representations to the interaction modeling results. We use the same number of units ($4d$) for each hidden layer.

Table \ref{tab:diff_layer} shows the results of using different MLP layers in GMCF. We can see that using the MLP with 0 hidden layer (GMCF-0) gains much worse results than other settings. It shows that a powerful non-linear algorithm helps extract useful information from inner interactions than a linear transformation. This result is consistent with the results in section \ref{sec:eval_inner_cross}. When the number of the hidden layer is 1, GMCF gains the best performance. This illustrates that deeper MLP does not necessarily increase the performance due to the overfitting \cite{he2017neural,guo2017deepfm}, and one hidden layer is enough in our models to analyze the inner interactions.

\begin{table}[t]
\caption{The performance of GMCF on using different MLP depths.}
\label{tab:diff_layer}
\footnotesize
\begin{center}
\vskip -0.1in
\begin{tabular}{l|>{\centering\arraybackslash}p{0.7cm}>{\centering\arraybackslash}p{1cm}|>{\centering\arraybackslash}p{0.7cm}>{\centering\arraybackslash}p{1cm}|>{\centering\arraybackslash}p{0.7cm}>{\centering\arraybackslash}p{1cm}}
\hline
 & \multicolumn{2}{c|}{\textbf{MovieLens 1M}} & \multicolumn{2}{c|}{\textbf{Book-Crossing}} & \multicolumn{2}{c}{\textbf{Taobao}}\\
 & AUC & NDCG@10 & AUC & NDCG@10 & AUC & NDCG@10\\
\hline
GMCF-0 & 0.9025 & 0.9354 & 0.8051 & 0.8869 & 0.6608 & 0.1341\\
GMCF-1 & \textbf{0.9127} & \textbf{0.9436} & \textbf{0.8228} & \textbf{0.8951} & \textbf{0.6679} & \textbf{0.1467}\\
GMCF-2 & 0.9109 & 0.9429 & 0.8182 & 0.8924 & 0.6662 & 0.1429\\
GMCF-3 & 0.9080 & 0.9411 & 0.8178 & 0.8905 & 0.6645 & 0.1395\\
GMCF-4 & 0.9071 & 0.9377 & 0.8153 & 0.8907 & 0.6650 & 0.1393\\
\hline
\end{tabular}
\end{center}
\vskip -0.2in
\end{table}

\vskip -0.18in
\subsection{Case Study}
\label{subsec:case_study}

In this section, we conduct case studies to evaluate whether our model learns collaborative information between attributes and whether the attributes show semantic meaning that provide potential explanations of the predictions. Specifically, we use the learned attribute embeddings from the MovieLens 1M dataset. We first calculate the embeddings' cosine similarities between user age groups (e.g., 1-18, 19-24) and between movie genres. Then, we calculate the node matching results between the the two types of attributes.
Figure \ref{fig:case_study} shows the similarity results (the left and central figures) and the node matching results (the right figure). Note that the darker the color, the higher the similarity or the node matching value. The age labels indicate the age groups (e.g., \textit{age\_18} means age 18-24).

From the left and central figures, we can see that similar user attributes and similar item attributes have similar embeddings after training. For example, in the left figure, the age group attributes are clearly divided into two groups at the age of 45, which means that the similar age attributes are grouped. It indicates that the same age group users may have a similar preference for movies in terms of age.
In the central figure, Fantasy, Sci-fi, and Mystery have similar embeddings in the latent space. Semantically, these genres are similar. Similar results are also shown for Crime, Horror, and Thriller. The above observations show that similar attributes are successfully learned to have similar embeddings in GMCF. 

The right figure shows the node matching between age groups and genres. We observe that different age groups have different preferences for movie genres. For example, younger users (under 45) have higher node matching values (preference) on fantasy, sci-fi, and mystery movies more than older users (over 45). In comparison, the older users seem to like musical and Documentary (\textit{Doc}) movies more. These observations show our model's ability to provide potential explanations about the prediction results at attribute level, e.g., we recommend \textit{Interstellar} (a sci-fi movie) to a 16-year-old user because younger people have a high chance to prefer a sci-fi movie.

\vspace{-0.8mm}
\section{Conclusion}

User and item attribute interactions provide useful information in recommender systems for accurate predictions. While existing work treat all the attribute interactions equally, in this work, we identify two types of attribute interactions: inner interactions and cross interactions. We propose a neural Graph Matching based Collaborative Filtering (GMCF) model. The GMCF model exploits the two types of interactions for different purpose in a graph matching structure. Specifically, GMCF explicitly performs user and item characteristic learning with the inner interactions, and performs preference matching for recommendation based on the cross interactions. Experimental results show that our GMCF model is effective and outperforms all state-of-the-art baselines in terms of accuracy on three widely used datasets. In future work, we will consider higher-order interactions in our model architecture, which may contain edge matching and sub-graph exploration.

\vskip -0.1in
\begin{acks}
This work is supported by the China Scholarship Council (CSC).
In this research, Junhao Gan was in part supported by Australian Research Council (ARC)
Discovery Early Career Researcher Award (DECRA) DE190101118. 
\end{acks}

\balance
\bibliography{reference}


\begin{thebibliography}{45}


\ifx \showCODEN    \undefined \def \showCODEN     #1{\unskip}     \fi
\ifx \showDOI      \undefined \def \showDOI       #1{#1}\fi
\ifx \showISBNx    \undefined \def \showISBNx     #1{\unskip}     \fi
\ifx \showISBNxiii \undefined \def \showISBNxiii  #1{\unskip}     \fi
\ifx \showISSN     \undefined \def \showISSN      #1{\unskip}     \fi
\ifx \showLCCN     \undefined \def \showLCCN      #1{\unskip}     \fi
\ifx \shownote     \undefined \def \shownote      #1{#1}          \fi
\ifx \showarticletitle \undefined \def \showarticletitle #1{#1}   \fi
\ifx \showURL      \undefined \def \showURL       {\relax}        \fi
\providecommand\bibfield[2]{#2}
\providecommand\bibinfo[2]{#2}
\providecommand\natexlab[1]{#1}
\providecommand\showeprint[2][]{arXiv:#2}

\bibitem[\protect\citeauthoryear{Adams, Dahl, and Murray}{Adams
  et~al\mbox{.}}{2010}]%
        {adams2010incorporating}
\bibfield{author}{\bibinfo{person}{Ryan~Prescott Adams},
  \bibinfo{person}{George~E Dahl}, {and} \bibinfo{person}{Iain Murray}.}
  \bibinfo{year}{2010}\natexlab{}.
\newblock \showarticletitle{Incorporating Side Information in Probabilistic
  Matrix Factorization with Gaussian Processes}. In
  \bibinfo{booktitle}{\emph{Proceedings of the Twenty-Sixth Conference on
  Uncertainty in Artificial Intelligence (UAI)}}. \bibinfo{pages}{1--9}.
\newblock


\bibitem[\protect\citeauthoryear{Bai, Ding, Bian, Chen, Sun, and Wang}{Bai
  et~al\mbox{.}}{2019}]%
        {bai2019simgnn}
\bibfield{author}{\bibinfo{person}{Yunsheng Bai}, \bibinfo{person}{Hao Ding},
  \bibinfo{person}{Song Bian}, \bibinfo{person}{Ting Chen},
  \bibinfo{person}{Yizhou Sun}, {and} \bibinfo{person}{Wei Wang}.}
  \bibinfo{year}{2019}\natexlab{}.
\newblock \showarticletitle{SimGNN: A Neural Network Approach to Fast Graph
  Similarity Computation}. In \bibinfo{booktitle}{\emph{Proceedings of the 35th
  International Conference on Web Search and Data Mining (WSDM)}}.
  \bibinfo{pages}{384--392}.
\newblock


\bibitem[\protect\citeauthoryear{Battaglia, Pascanu, Lai, Rezende, and
  Kavukcuoglu}{Battaglia et~al\mbox{.}}{2016}]%
        {battaglia2016interaction}
\bibfield{author}{\bibinfo{person}{Peter~W. Battaglia}, \bibinfo{person}{Razvan
  Pascanu}, \bibinfo{person}{Matthew Lai}, \bibinfo{person}{Danilo~Jimenez
  Rezende}, {and} \bibinfo{person}{Koray Kavukcuoglu}.}
  \bibinfo{year}{2016}\natexlab{}.
\newblock \showarticletitle{Interaction Networks for Learning about Objects,
  Relations and Physics}. In \bibinfo{booktitle}{\emph{Proceedings of the
  Advances in Neural Information Processing Systems (NeurIPS)}}.
  \bibinfo{pages}{4502--4510}.
\newblock


\bibitem[\protect\citeauthoryear{Beutel, Covington, Jain, Xu, Li, Gatto, and
  Chi}{Beutel et~al\mbox{.}}{2018}]%
        {beutel2018latent}
\bibfield{author}{\bibinfo{person}{Alex Beutel}, \bibinfo{person}{Paul
  Covington}, \bibinfo{person}{Sagar Jain}, \bibinfo{person}{Can Xu},
  \bibinfo{person}{Jia Li}, \bibinfo{person}{Vince Gatto}, {and}
  \bibinfo{person}{Ed~H Chi}.} \bibinfo{year}{2018}\natexlab{}.
\newblock \showarticletitle{Latent Cross: Making Use of Context in Recurrent
  Recommender Systems}. In \bibinfo{booktitle}{\emph{Proceedings of the 11th
  ACM International Conference on Web Search and Data Mining (WSDM)}}.
  \bibinfo{pages}{46--54}.
\newblock


\bibitem[\protect\citeauthoryear{Bromley, Guyon, LeCun, S{\"a}ckinger, and
  Shah}{Bromley et~al\mbox{.}}{1994}]%
        {bromley1994signature}
\bibfield{author}{\bibinfo{person}{Jane Bromley}, \bibinfo{person}{Isabelle
  Guyon}, \bibinfo{person}{Yann LeCun}, \bibinfo{person}{Eduard S{\"a}ckinger},
  {and} \bibinfo{person}{Roopak Shah}.} \bibinfo{year}{1994}\natexlab{}.
\newblock \showarticletitle{Signature Verification Using a "Siamese" Time Delay
  Neural Network}. In \bibinfo{booktitle}{\emph{Proceedings of the Advances in
  Neural Information Processing Systems (NeurIPS)}}. \bibinfo{pages}{737--744}.
\newblock


\bibitem[\protect\citeauthoryear{Chang, Ullman, Torralba, and Tenenbaum}{Chang
  et~al\mbox{.}}{2016}]%
        {chang2016compositional}
\bibfield{author}{\bibinfo{person}{Michael~B Chang}, \bibinfo{person}{Tomer
  Ullman}, \bibinfo{person}{Antonio Torralba}, {and} \bibinfo{person}{Joshua~B
  Tenenbaum}.} \bibinfo{year}{2016}\natexlab{}.
\newblock \showarticletitle{A Compositional Object-based Approach to Learning
  Physical Dynamics}. In \bibinfo{booktitle}{\emph{Proceedings of the 5th
  International Conference on Learning Representations (ICLR)}}.
  \bibinfo{pages}{1--15}.
\newblock


\bibitem[\protect\citeauthoryear{Cheng, Koc, Harmsen, Shaked, Chandra, Aradhye,
  Anderson, Corrado, Chai, Ispir, Anil, Haque, Hong, Jain, Liu, and Shah}{Cheng
  et~al\mbox{.}}{2016}]%
        {cheng2016wide}
\bibfield{author}{\bibinfo{person}{Heng{-}Tze Cheng}, \bibinfo{person}{Levent
  Koc}, \bibinfo{person}{Jeremiah Harmsen}, \bibinfo{person}{Tal Shaked},
  \bibinfo{person}{Tushar Chandra}, \bibinfo{person}{Hrishi Aradhye},
  \bibinfo{person}{Glen Anderson}, \bibinfo{person}{Greg Corrado},
  \bibinfo{person}{Wei Chai}, \bibinfo{person}{Mustafa Ispir},
  \bibinfo{person}{Rohan Anil}, \bibinfo{person}{Zakaria Haque},
  \bibinfo{person}{Lichan Hong}, \bibinfo{person}{Vihan Jain},
  \bibinfo{person}{Xiaobing Liu}, {and} \bibinfo{person}{Hemal Shah}.}
  \bibinfo{year}{2016}\natexlab{}.
\newblock \showarticletitle{Wide \& Deep Learning for Recommender Systems}. In
  \bibinfo{booktitle}{\emph{Proceedings of the 1st Workshop on Deep Learning
  for Recommender Systems (RecSys)}}. \bibinfo{pages}{7--10}.
\newblock


\bibitem[\protect\citeauthoryear{Dijkman, Dumas, and
  Garc{\'\i}a-Ba{\~n}uelos}{Dijkman et~al\mbox{.}}{2009}]%
        {dijkman2009graph}
\bibfield{author}{\bibinfo{person}{Remco Dijkman}, \bibinfo{person}{Marlon
  Dumas}, {and} \bibinfo{person}{Luciano Garc{\'\i}a-Ba{\~n}uelos}.}
  \bibinfo{year}{2009}\natexlab{}.
\newblock \showarticletitle{Graph Matching Algorithms for Business Process
  Model Similarity Search}. In \bibinfo{booktitle}{\emph{Proceedings of the 7th
  International Conference on Business Process Management (BPM)}}. Springer,
  \bibinfo{pages}{48--63}.
\newblock


\bibitem[\protect\citeauthoryear{Duvenaud, Maclaurin, Iparraguirre, Bombarell,
  Hirzel, Aspuru-Guzik, and Adams}{Duvenaud et~al\mbox{.}}{2015}]%
        {duvenaud2015convolutional}
\bibfield{author}{\bibinfo{person}{David~K Duvenaud}, \bibinfo{person}{Dougal
  Maclaurin}, \bibinfo{person}{Jorge Iparraguirre}, \bibinfo{person}{Rafael
  Bombarell}, \bibinfo{person}{Timothy Hirzel}, \bibinfo{person}{Al{\'a}n
  Aspuru-Guzik}, {and} \bibinfo{person}{Ryan~P Adams}.}
  \bibinfo{year}{2015}\natexlab{}.
\newblock \showarticletitle{Convolutional networks on graphs for learning
  molecular fingerprints}. In \bibinfo{booktitle}{\emph{Proceedings of the
  Advances in Neural Information Processing Systems (NeurIPS)}}.
  \bibinfo{pages}{2224--2232}.
\newblock


\bibitem[\protect\citeauthoryear{Gilmer, Schoenholz, Riley, Vinyals, and
  Dahl}{Gilmer et~al\mbox{.}}{2017}]%
        {gilmer2017neural}
\bibfield{author}{\bibinfo{person}{Justin Gilmer}, \bibinfo{person}{Samuel~S
  Schoenholz}, \bibinfo{person}{Patrick~F Riley}, \bibinfo{person}{Oriol
  Vinyals}, {and} \bibinfo{person}{George~E Dahl}.}
  \bibinfo{year}{2017}\natexlab{}.
\newblock \showarticletitle{Neural Message Passing for Quantum Chemistry}. In
  \bibinfo{booktitle}{\emph{Proceedings of the 34th International Conference on
  Machine Learning (ICML)}}. \bibinfo{pages}{1263--1272}.
\newblock


\bibitem[\protect\citeauthoryear{Guo, Tang, Ye, Li, and He}{Guo
  et~al\mbox{.}}{2017}]%
        {guo2017deepfm}
\bibfield{author}{\bibinfo{person}{Huifeng Guo}, \bibinfo{person}{Ruiming
  Tang}, \bibinfo{person}{Yunming Ye}, \bibinfo{person}{Zhenguo Li}, {and}
  \bibinfo{person}{Xiuqiang He}.} \bibinfo{year}{2017}\natexlab{}.
\newblock \showarticletitle{DeepFM: a Factorization-Machine based Neural
  Network for CTR prediction}. In \bibinfo{booktitle}{\emph{Proceedings of the
  26th International Joint Conference on Artificial Intelligence (IJCAI)}}.
  \bibinfo{pages}{1725--1731}.
\newblock


\bibitem[\protect\citeauthoryear{Harper and Konstan}{Harper and
  Konstan}{2015}]%
        {harper2015movielens}
\bibfield{author}{\bibinfo{person}{F~Maxwell Harper} {and}
  \bibinfo{person}{Joseph~A Konstan}.} \bibinfo{year}{2015}\natexlab{}.
\newblock \showarticletitle{The Movielens Datasets: History and Context}.
\newblock \bibinfo{journal}{\emph{Transactions on Interactive Intelligent
  Systems (TIIS)}} (\bibinfo{year}{2015}), \bibinfo{pages}{1--19}.
\newblock


\bibitem[\protect\citeauthoryear{He and Chua}{He and Chua}{2017}]%
        {he2017neural}
\bibfield{author}{\bibinfo{person}{Xiangnan He} {and} \bibinfo{person}{Tat-Seng
  Chua}.} \bibinfo{year}{2017}\natexlab{}.
\newblock \showarticletitle{Neural Factorization Machines for Sparse Predictive
  Analytics}. In \bibinfo{booktitle}{\emph{Proceedings of the 40th
  International ACM conference on Research and Development in Information
  Retrieval (SIGIR)}}. \bibinfo{pages}{355--364}.
\newblock


\bibitem[\protect\citeauthoryear{He, Liao, Zhang, Nie, Hu, and Chua}{He
  et~al\mbox{.}}{2017}]%
        {he2017neuralCF}
\bibfield{author}{\bibinfo{person}{Xiangnan He}, \bibinfo{person}{Lizi Liao},
  \bibinfo{person}{Hanwang Zhang}, \bibinfo{person}{Liqiang Nie},
  \bibinfo{person}{Xia Hu}, {and} \bibinfo{person}{Tat-Seng Chua}.}
  \bibinfo{year}{2017}\natexlab{}.
\newblock \showarticletitle{Neural Collaborative Filtering}. In
  \bibinfo{booktitle}{\emph{Proceedings of the 26th international conference on
  world wide web (WWW)}}. \bibinfo{pages}{173--182}.
\newblock


\bibitem[\protect\citeauthoryear{Huang, Qi, Sun, Zhang, and Zheng}{Huang
  et~al\mbox{.}}{2019}]%
        {huang2019carl}
\bibfield{author}{\bibinfo{person}{Xinting Huang}, \bibinfo{person}{Jianzhong
  Qi}, \bibinfo{person}{Yu Sun}, \bibinfo{person}{Rui Zhang}, {and}
  \bibinfo{person}{Hai-Tao Zheng}.} \bibinfo{year}{2019}\natexlab{}.
\newblock \showarticletitle{CARL: Aggregated Search with Context-Aware Module
  Embedding Learning}. In \bibinfo{booktitle}{\emph{International Joint
  Conference on Neural Networks (IJCNN)}}. IEEE, \bibinfo{pages}{101--108}.
\newblock


\bibitem[\protect\citeauthoryear{Kashima, Tsuda, and Inokuchi}{Kashima
  et~al\mbox{.}}{2003}]%
        {kashima2003marginalized}
\bibfield{author}{\bibinfo{person}{Hisashi Kashima}, \bibinfo{person}{Koji
  Tsuda}, {and} \bibinfo{person}{Akihiro Inokuchi}.}
  \bibinfo{year}{2003}\natexlab{}.
\newblock \showarticletitle{Marginalized kernels between labeled graphs}. In
  \bibinfo{booktitle}{\emph{Proceedings of the 20th International Conference on
  Machine Learning (ICML)}}. \bibinfo{pages}{321--328}.
\newblock


\bibitem[\protect\citeauthoryear{Kingma and Ba}{Kingma and Ba}{2015}]%
        {kingma2014adam}
\bibfield{author}{\bibinfo{person}{Diederik~P Kingma} {and}
  \bibinfo{person}{Jimmy Ba}.} \bibinfo{year}{2015}\natexlab{}.
\newblock \showarticletitle{Adam: A Method for Stochastic Optimization}. In
  \bibinfo{booktitle}{\emph{Proceedings of the 4th International Conference on
  Learning Representations (ICLR)}}. \bibinfo{pages}{1--15}.
\newblock


\bibitem[\protect\citeauthoryear{Kipf and Welling}{Kipf and Welling}{2017}]%
        {kipf2016semi}
\bibfield{author}{\bibinfo{person}{Thomas~N Kipf} {and} \bibinfo{person}{Max
  Welling}.} \bibinfo{year}{2017}\natexlab{}.
\newblock \showarticletitle{Semi-supervised classification with graph
  convolutional networks}. In \bibinfo{booktitle}{\emph{Proceedings of the 6th
  International Conference on Learning Representations (ICLR)}}.
  \bibinfo{pages}{1--14}.
\newblock


\bibitem[\protect\citeauthoryear{Koren, Bell, and Volinsky}{Koren
  et~al\mbox{.}}{2009}]%
        {koren2009matrix}
\bibfield{author}{\bibinfo{person}{Yehuda Koren}, \bibinfo{person}{Robert
  Bell}, {and} \bibinfo{person}{Chris Volinsky}.}
  \bibinfo{year}{2009}\natexlab{}.
\newblock \showarticletitle{Matrix Factorization Techniques for Recommender
  Systems}.
\newblock \bibinfo{journal}{\emph{Computer}} \bibinfo{volume}{42},
  \bibinfo{number}{8} (\bibinfo{year}{2009}), \bibinfo{pages}{30--37}.
\newblock


\bibitem[\protect\citeauthoryear{Li, Gu, Dullien, Vinyals, and Kohli}{Li
  et~al\mbox{.}}{2019b}]%
        {li2019graph}
\bibfield{author}{\bibinfo{person}{Yujia Li}, \bibinfo{person}{Chenjie Gu},
  \bibinfo{person}{Thomas Dullien}, \bibinfo{person}{Oriol Vinyals}, {and}
  \bibinfo{person}{Pushmeet Kohli}.} \bibinfo{year}{2019}\natexlab{b}.
\newblock \showarticletitle{Graph Matching Networks for Learning the Similarity
  of Graph Structured Objects}. In \bibinfo{booktitle}{\emph{Proceedings of the
  36th International Conference on Machine Learning (ICML)}}.
  \bibinfo{pages}{3835--3845}.
\newblock


\bibitem[\protect\citeauthoryear{Li, Cui, Wu, Zhang, and Wang}{Li
  et~al\mbox{.}}{2019a}]%
        {li2019fi}
\bibfield{author}{\bibinfo{person}{Zekun Li}, \bibinfo{person}{Zeyu Cui},
  \bibinfo{person}{Shu Wu}, \bibinfo{person}{Xiaoyu Zhang}, {and}
  \bibinfo{person}{Liang Wang}.} \bibinfo{year}{2019}\natexlab{a}.
\newblock \showarticletitle{Fi-GNN: Modeling Feature Interactions via Graph
  Neural Networks for CTR Prediction}. In \bibinfo{booktitle}{\emph{Proceedings
  of the 28th International Conference on Information and Knowledge Management
  (CIKM)}}. \bibinfo{pages}{539--548}.
\newblock


\bibitem[\protect\citeauthoryear{Liu, Zhu, Li, Zhang, Lai, Tang, He, Li, and
  Yu}{Liu et~al\mbox{.}}{2020}]%
        {liu2020autofis}
\bibfield{author}{\bibinfo{person}{Bin Liu}, \bibinfo{person}{Chenxu Zhu},
  \bibinfo{person}{Guilin Li}, \bibinfo{person}{Weinan Zhang},
  \bibinfo{person}{Jincai Lai}, \bibinfo{person}{Ruiming Tang},
  \bibinfo{person}{Xiuqiang He}, \bibinfo{person}{Zhenguo Li}, {and}
  \bibinfo{person}{Yong Yu}.} \bibinfo{year}{2020}\natexlab{}.
\newblock \showarticletitle{AutoFIS: Automatic Feature Interaction Selection in
  Factorization Models for Click-Through Rate Prediction}. In
  \bibinfo{booktitle}{\emph{Proceedings of the 26th International Conference on
  Knowledge Discovery and Data Mining (SIGKDD)}}. \bibinfo{publisher}{ACM},
  \bibinfo{pages}{2636--2645}.
\newblock


\bibitem[\protect\citeauthoryear{Pang, Zhao, and Li}{Pang
  et~al\mbox{.}}{2021}]%
        {pang2021graph}
\bibfield{author}{\bibinfo{person}{Yunsheng Pang}, \bibinfo{person}{Yunxiang
  Zhao}, {and} \bibinfo{person}{Dongsheng Li}.}
  \bibinfo{year}{2021}\natexlab{}.
\newblock \showarticletitle{Graph Pooling via Coarsened Graph Infomax}.
\newblock \bibinfo{journal}{\emph{arXiv preprint arXiv:2105.01275}}
  (\bibinfo{year}{2021}).
\newblock


\bibitem[\protect\citeauthoryear{Raymond, Gardiner, and Willett}{Raymond
  et~al\mbox{.}}{2002}]%
        {raymond2002rascal}
\bibfield{author}{\bibinfo{person}{John~W Raymond}, \bibinfo{person}{Eleanor~J
  Gardiner}, {and} \bibinfo{person}{Peter Willett}.}
  \bibinfo{year}{2002}\natexlab{}.
\newblock \showarticletitle{Rascal: Calculation of Graph Similarity Using
  Maximum Common Edge Subgraphs}.
\newblock \bibinfo{journal}{\emph{Comput. J.}} \bibinfo{volume}{45},
  \bibinfo{number}{6} (\bibinfo{year}{2002}), \bibinfo{pages}{631--644}.
\newblock


\bibitem[\protect\citeauthoryear{Rendle}{Rendle}{2010}]%
        {rendle2010factorization}
\bibfield{author}{\bibinfo{person}{Steffen Rendle}.}
  \bibinfo{year}{2010}\natexlab{}.
\newblock \showarticletitle{Factorization Machines}. In
  \bibinfo{booktitle}{\emph{Proceedings of the 10th International IEEE
  Conference on Data Mining (ICDM)}}. \bibinfo{pages}{995--1000}.
\newblock


\bibitem[\protect\citeauthoryear{Shervashidze and Borgwardt}{Shervashidze and
  Borgwardt}{2009}]%
        {shervashidze2009fast}
\bibfield{author}{\bibinfo{person}{Nino Shervashidze} {and}
  \bibinfo{person}{Karsten Borgwardt}.} \bibinfo{year}{2009}\natexlab{}.
\newblock \showarticletitle{Fast Subtree Kernels on Graphs}. In
  \bibinfo{booktitle}{\emph{Proceedings of the Advances in Neural Information
  Processing Systems (NeurIPS)}}. \bibinfo{pages}{1660--1668}.
\newblock


\bibitem[\protect\citeauthoryear{Shervashidze, Vishwanathan, Petri, Mehlhorn,
  and Borgwardt}{Shervashidze et~al\mbox{.}}{2009}]%
        {shervashidze2009efficient}
\bibfield{author}{\bibinfo{person}{Nino Shervashidze}, \bibinfo{person}{SVN
  Vishwanathan}, \bibinfo{person}{Tobias Petri}, \bibinfo{person}{Kurt
  Mehlhorn}, {and} \bibinfo{person}{Karsten Borgwardt}.}
  \bibinfo{year}{2009}\natexlab{}.
\newblock \showarticletitle{Efficient Graphlet Kernels for Large Graph
  Comparison}. In \bibinfo{booktitle}{\emph{Proceedings of the 12th
  International Conference on Artificial Intelligence and Statistics
  (AISTATS)}}. \bibinfo{pages}{488--495}.
\newblock


\bibitem[\protect\citeauthoryear{Song, Shi, Xiao, Duan, Xu, Zhang, and
  Tang}{Song et~al\mbox{.}}{2019}]%
        {song2019autoint}
\bibfield{author}{\bibinfo{person}{Weiping Song}, \bibinfo{person}{Chence Shi},
  \bibinfo{person}{Zhiping Xiao}, \bibinfo{person}{Zhijian Duan},
  \bibinfo{person}{Yewen Xu}, \bibinfo{person}{Ming Zhang}, {and}
  \bibinfo{person}{Jian Tang}.} \bibinfo{year}{2019}\natexlab{}.
\newblock \showarticletitle{Autoint: Automatic Feature Interaction Learning via
  Self-attentive Neural Networks}. In \bibinfo{booktitle}{\emph{Proceedings of
  the 28th International Conference on Information and Knowledge Management
  (CIKM)}}. \bibinfo{pages}{1161--1170}.
\newblock


\bibitem[\protect\citeauthoryear{Su, Erfani, and Zhang}{Su
  et~al\mbox{.}}{2019}]%
        {su2019mmf}
\bibfield{author}{\bibinfo{person}{Yixin Su}, \bibinfo{person}{Sarah~Monazam
  Erfani}, {and} \bibinfo{person}{Rui Zhang}.} \bibinfo{year}{2019}\natexlab{}.
\newblock \showarticletitle{MMF: Attribute Interpretable Collaborative
  Filtering}. In \bibinfo{booktitle}{\emph{International Joint Conference on
  Neural Networks (IJCNN)}}. IEEE, \bibinfo{pages}{1--8}.
\newblock


\bibitem[\protect\citeauthoryear{Su, Zhang, Erfani, and Xu}{Su
  et~al\mbox{.}}{2021}]%
        {su2020detecting}
\bibfield{author}{\bibinfo{person}{Yixin Su}, \bibinfo{person}{Rui Zhang},
  \bibinfo{person}{Sarah Erfani}, {and} \bibinfo{person}{Zhenghua Xu}.}
  \bibinfo{year}{2021}\natexlab{}.
\newblock \showarticletitle{Detecting Beneficial Feature Interactions for
  Recommender Systems}. In \bibinfo{booktitle}{\emph{Proceedings of the
  Conference on Artificial Intelligence (AAAI)}}.
\newblock


\bibitem[\protect\citeauthoryear{van~den Berg, Kipf, and Welling}{van~den Berg
  et~al\mbox{.}}{2017}]%
        {berg2017graph}
\bibfield{author}{\bibinfo{person}{Rianne van~den Berg},
  \bibinfo{person}{Thomas~N. Kipf}, {and} \bibinfo{person}{Max Welling}.}
  \bibinfo{year}{2017}\natexlab{}.
\newblock \bibinfo{title}{Graph Convolutional Matrix Completion}.
\newblock
\newblock
\showeprint[arxiv]{1706.02263}~[stat.ML]


\bibitem[\protect\citeauthoryear{Vishwanathan, Schraudolph, Kondor, and
  Borgwardt}{Vishwanathan et~al\mbox{.}}{2010}]%
        {vishwanathan2010graph}
\bibfield{author}{\bibinfo{person}{S~Vichy~N Vishwanathan},
  \bibinfo{person}{Nicol~N Schraudolph}, \bibinfo{person}{Risi Kondor}, {and}
  \bibinfo{person}{Karsten~M Borgwardt}.} \bibinfo{year}{2010}\natexlab{}.
\newblock \showarticletitle{Graph Kernels}.
\newblock \bibinfo{journal}{\emph{The Journal of Machine Learning Research
  (JMLR)}}  \bibinfo{volume}{11} (\bibinfo{year}{2010}),
  \bibinfo{pages}{1201--1242}.
\newblock


\bibitem[\protect\citeauthoryear{Wang, Zhang, Zhang, Leskovec, Zhao, Li, and
  Wang}{Wang et~al\mbox{.}}{2019c}]%
        {wang2019knowledge}
\bibfield{author}{\bibinfo{person}{Hongwei Wang}, \bibinfo{person}{Fuzheng
  Zhang}, \bibinfo{person}{Mengdi Zhang}, \bibinfo{person}{Jure Leskovec},
  \bibinfo{person}{Miao Zhao}, \bibinfo{person}{Wenjie Li}, {and}
  \bibinfo{person}{Zhongyuan Wang}.} \bibinfo{year}{2019}\natexlab{c}.
\newblock \showarticletitle{Knowledge-Aware Graph Neural Networks with Label
  Smoothness Regularization for Recommender Systems}. In
  \bibinfo{booktitle}{\emph{Proceedings of the 25th International Conference on
  Knowledge Discovery and Data Mining (SIGKDD)}}. \bibinfo{pages}{968--977}.
\newblock


\bibitem[\protect\citeauthoryear{Wang, He, Cao, Liu, and Chua}{Wang
  et~al\mbox{.}}{2019a}]%
        {wang2019kgat}
\bibfield{author}{\bibinfo{person}{Xiang Wang}, \bibinfo{person}{Xiangnan He},
  \bibinfo{person}{Yixin Cao}, \bibinfo{person}{Meng Liu}, {and}
  \bibinfo{person}{Tat-Seng Chua}.} \bibinfo{year}{2019}\natexlab{a}.
\newblock \showarticletitle{Kgat: Knowledge Graph Attention Network for
  Recommendation}. In \bibinfo{booktitle}{\emph{Proceedings of the 25th
  International Conference on Knowledge Discovery and Data Mining (SIGKDD)}}.
  \bibinfo{pages}{950--958}.
\newblock


\bibitem[\protect\citeauthoryear{Wang, He, Wang, Feng, and Chua}{Wang
  et~al\mbox{.}}{2019b}]%
        {wang2019neural}
\bibfield{author}{\bibinfo{person}{Xiang Wang}, \bibinfo{person}{Xiangnan He},
  \bibinfo{person}{Meng Wang}, \bibinfo{person}{Fuli Feng}, {and}
  \bibinfo{person}{Tat-Seng Chua}.} \bibinfo{year}{2019}\natexlab{b}.
\newblock \showarticletitle{Neural Graph Collaborative Filtering}. In
  \bibinfo{booktitle}{\emph{Proceedings of the 42nd International conference on
  Research and Development in Information Retrieval (SIGIR)}}.
  \bibinfo{pages}{165--174}.
\newblock


\bibitem[\protect\citeauthoryear{Wang, Zhang, Sun, and Qi}{Wang
  et~al\mbox{.}}{2021}]%
        {wang2021combating}
\bibfield{author}{\bibinfo{person}{Xiaojie Wang}, \bibinfo{person}{Rui Zhang},
  \bibinfo{person}{Yu Sun}, {and} \bibinfo{person}{Jianzhong Qi}.}
  \bibinfo{year}{2021}\natexlab{}.
\newblock \showarticletitle{Combating Selection Biases in Recommender Systems
  with a Few Unbiased Ratings}. In \bibinfo{booktitle}{\emph{Proceedings of the
  14th ACM International Conference on Web Search and Data Mining (WSDM)}}.
  \bibinfo{pages}{427--435}.
\newblock


\bibitem[\protect\citeauthoryear{Willett, Barnard, and Downs}{Willett
  et~al\mbox{.}}{1998}]%
        {willett1998chemical}
\bibfield{author}{\bibinfo{person}{Peter Willett}, \bibinfo{person}{John~M
  Barnard}, {and} \bibinfo{person}{Geoffrey~M Downs}.}
  \bibinfo{year}{1998}\natexlab{}.
\newblock \showarticletitle{Chemical Similarity Searching}.
\newblock \bibinfo{journal}{\emph{Chemical Information and Computer Sciences}}
  (\bibinfo{year}{1998}), \bibinfo{pages}{983--996}.
\newblock


\bibitem[\protect\citeauthoryear{Xian, Fu, Muthukrishnan, De~Melo, and
  Zhang}{Xian et~al\mbox{.}}{2019}]%
        {xian2019reinforcement}
\bibfield{author}{\bibinfo{person}{Yikun Xian}, \bibinfo{person}{Zuohui Fu},
  \bibinfo{person}{S Muthukrishnan}, \bibinfo{person}{Gerard De~Melo}, {and}
  \bibinfo{person}{Yongfeng Zhang}.} \bibinfo{year}{2019}\natexlab{}.
\newblock \showarticletitle{Reinforcement Knowledge Graph Reasoning for
  Explainable Recommendation}. In \bibinfo{booktitle}{\emph{Proceedings of the
  42nd International Conference on Research and Development in Information
  Retrieval (SIGIR)}}. \bibinfo{pages}{285--294}.
\newblock


\bibitem[\protect\citeauthoryear{Xiao, Ye, He, Zhang, Wu, and Chua}{Xiao
  et~al\mbox{.}}{2017}]%
        {xiao2017attentional}
\bibfield{author}{\bibinfo{person}{Jun Xiao}, \bibinfo{person}{Hao Ye},
  \bibinfo{person}{Xiangnan He}, \bibinfo{person}{Hanwang Zhang},
  \bibinfo{person}{Fei Wu}, {and} \bibinfo{person}{Tat-Seng Chua}.}
  \bibinfo{year}{2017}\natexlab{}.
\newblock \showarticletitle{Attentional Factorization Machines: Learning the
  Weight of Feature Interactions via Attention Networks}. In
  \bibinfo{booktitle}{\emph{Proceedings of the 26th International Joint
  Conference on Artificial Intelligence (IJCAI)}}. \bibinfo{pages}{3119--3125}.
\newblock


\bibitem[\protect\citeauthoryear{Xu, Hu, Leskovec, and Jegelka}{Xu
  et~al\mbox{.}}{2019}]%
        {xu2018powerful}
\bibfield{author}{\bibinfo{person}{Keyulu Xu}, \bibinfo{person}{Weihua Hu},
  \bibinfo{person}{Jure Leskovec}, {and} \bibinfo{person}{Stefanie Jegelka}.}
  \bibinfo{year}{2019}\natexlab{}.
\newblock \showarticletitle{How Powerful are Graph Neural Networks?}. In
  \bibinfo{booktitle}{\emph{Proceedings of the 8th International Conference on
  Learning Representations (ICLR)}}. \bibinfo{pages}{1--17}.
\newblock


\bibitem[\protect\citeauthoryear{Yan, Yu, and Han}{Yan et~al\mbox{.}}{2005}]%
        {yan2005substructure}
\bibfield{author}{\bibinfo{person}{Xifeng Yan}, \bibinfo{person}{Philip~S Yu},
  {and} \bibinfo{person}{Jiawei Han}.} \bibinfo{year}{2005}\natexlab{}.
\newblock \showarticletitle{Substructure Similarity Search in Graph Databases}.
  In \bibinfo{booktitle}{\emph{Proceedings of the 2005 International Conference
  on Management of Data (SIGMOD)}}. \bibinfo{pages}{766--777}.
\newblock


\bibitem[\protect\citeauthoryear{Zhang, Yuan, Lian, Xie, and Ma}{Zhang
  et~al\mbox{.}}{2016}]%
        {zhang2016collaborative}
\bibfield{author}{\bibinfo{person}{Fuzheng Zhang},
  \bibinfo{person}{Nicholas~Jing Yuan}, \bibinfo{person}{Defu Lian},
  \bibinfo{person}{Xing Xie}, {and} \bibinfo{person}{Wei-Ying Ma}.}
  \bibinfo{year}{2016}\natexlab{}.
\newblock \showarticletitle{Collaborative Knowledge base Embedding for
  Recommender Systems}. In \bibinfo{booktitle}{\emph{Proceedings of the 22nd
  international conference on knowledge discovery and data mining (SIGKDD)}}.
  \bibinfo{pages}{353--362}.
\newblock


\bibitem[\protect\citeauthoryear{Zhao, Qi, Liu, and Zhang}{Zhao
  et~al\mbox{.}}{2021}]%
        {zhao2021wgcn}
\bibfield{author}{\bibinfo{person}{Yunxiang Zhao}, \bibinfo{person}{Jianzhong
  Qi}, \bibinfo{person}{Qingwei Liu}, {and} \bibinfo{person}{Rui Zhang}.}
  \bibinfo{year}{2021}\natexlab{}.
\newblock \showarticletitle{WGCN: Graph Convolutional Networks with Weighted
  Structural Features}.
\newblock \bibinfo{journal}{\emph{arXiv preprint arXiv:2104.14060}}
  (\bibinfo{year}{2021}).
\newblock


\bibitem[\protect\citeauthoryear{Zhou, Zhu, Song, Fan, Zhu, Ma, Yan, Jin, Li,
  and Gai}{Zhou et~al\mbox{.}}{2018}]%
        {zhou2018deep}
\bibfield{author}{\bibinfo{person}{Guorui Zhou}, \bibinfo{person}{Xiaoqiang
  Zhu}, \bibinfo{person}{Chenru Song}, \bibinfo{person}{Ying Fan},
  \bibinfo{person}{Han Zhu}, \bibinfo{person}{Xiao Ma},
  \bibinfo{person}{Yanghui Yan}, \bibinfo{person}{Junqi Jin},
  \bibinfo{person}{Han Li}, {and} \bibinfo{person}{Kun Gai}.}
  \bibinfo{year}{2018}\natexlab{}.
\newblock \showarticletitle{Deep Interest Network for Click-through Rate
  Prediction}. In \bibinfo{booktitle}{\emph{Proceedings of the 24th ACM
  International Conference on Knowledge Discovery and Data Mining (SIGKDD)}}.
  \bibinfo{pages}{1059--1068}.
\newblock


\bibitem[\protect\citeauthoryear{Ziegler, McNee, Konstan, and Lausen}{Ziegler
  et~al\mbox{.}}{2005}]%
        {ziegler2005improving}
\bibfield{author}{\bibinfo{person}{Cai-Nicolas Ziegler},
  \bibinfo{person}{Sean~M McNee}, \bibinfo{person}{Joseph~A Konstan}, {and}
  \bibinfo{person}{Georg Lausen}.} \bibinfo{year}{2005}\natexlab{}.
\newblock \showarticletitle{Improving Recommendation Lists Through Topic
  Diversification}. In \bibinfo{booktitle}{\emph{Proceedings of the 14th
  International Conference on World Wide Web (WWW)}}. \bibinfo{pages}{22--32}.
\newblock


\end{thebibliography}
\bibliographystyle{ACM-Reference-Format}

\end{document}